\documentclass[aps,prb,twocolumn,showpacs,floatfix]{revtex4}
\usepackage{amssymb}
\usepackage{graphicx}
\usepackage{dcolumn}
\usepackage{bm}
\usepackage{amsmath}

\graphicspath{{figs_v3/}}

\begin{document}

\title{Interplay between magnetic properties and Fermi surface nesting in
iron pnictides}
\author{A.~N. Yaresko}
\author{G.-Q. Liu}
\author{V.~N. Antonov}
\author{O.~K. Andersen}
\affiliation{Max-Planck-Institut f\"ur Festk\"orperforschung, Heisenbergstra{\ss }e 1,
D-70569 Stuttgart, Germany}
\date{\today }

\begin{abstract}
The wave-vector $\left( \mathbf{q}\right) $ and doping $(x,y)$ dependences
of the magnetic energy, iron moment, and effective exchange interactions in
LaFeAsO$_{1-x}$F$_{x}$ and Ba$_{1-2y}$K$_{2y}$Fe$_{2}$As$_{2}$ are studied
by self-consistent LSDA calculations for co-planar spin spirals. For the
undoped compounds $\left( x=0,\,y=0\right) $, the minimum of the calculated
total energy, $E(\mathbf{q})$, is for $\mathbf{q}$ corresponding to stripe
antiferromagnetic order. Already at low levels of electron doping $(x)$,
this minimum becomes flat in LaFeAsO$_{1-x}$F$_{x}$ and for $x\gtrsim 5\%$, 
it shifts to an incommensurate $\mathbf{q}$. In Ba$_{1-2y}$K$_{2y}$Fe$%
_{2}$As$_{2}$, stripe order remains stable for hole doping up to $y=0.3$.
These results are explained in terms of the band structure. The magnetic
interactions cannot be accurately described by a simple classical Heisenberg
model and the effective exchange interactions fitted to $E(\mathbf{q})$\
depend strongly on doping. The doping dependence of the $E(\mathbf{q})$\
curves is compared with that of the noninteracting magnetic susceptibility
for which similar trends are found.
\end{abstract}

\maketitle

\section{\label{sec:intro}Introduction}

The discovery of superconductivity with $T_{c}$=27 K in F-doped LaFeAsO$%
_{1-x}$F$_{x}$ by Hosono and co-workers \cite{KWHH08} a year ago initiated
an avalanche of experimental and theoretical investigations of layered iron
pnictides and recently also chalcogenides. Soon, the superconducting
transition temperature was raised above 50 K by substituting La by smaller
rare-earth ions. \cite{RYLY+08_Pr,RYLY+08_Nd} The interest in layered iron
pnictides increased even more when superconductivity below $T_{c}$=38 K was
reported in oxygen-free potassium-doped Ba$_{1-2y}$K$_{2y}$Fe$_{2}$As$_{2}$,
\cite{RTJ08} for which good quality single crystals could be synthesized. 
\cite{NBKN+08}

Both families of iron pnictides have a quasi-two-dimensional (2D) tetragonal
crystal structure, in which FeAs layers are separated by either LaO or Ba
layers. The Fe ions form a square lattice sandwiched between two As sheets
shifted so that each Fe is surrounded by a slightly squeezed As tetrahedron.
At about 150 K, both stoichiometric parent compounds undergo a structural
transition at which the symmetry of the lattice lowers to orthorhombic.\cite%
{NKKH+08,RTJS+08} Magnetic order sets in at the same temperature as the
structural transition in BaFe$_{2}$As$_{2}$, but at a 20 K lower
temperature in LaFeAsO. In both cases, the order is striped: ferromagnetic
(FM) along the shorter axis of the square Fe sublattice and
antiferromagnetic (AFM) along the longer axis and between the Fe layers.\cite%
{RTJS+08,x:GABB+08,HQBG+08} The Fe moments are 0.4--0.9 $\mu _{\text{B}}$
in BaFe$_{2}$As$_{2}$ and 0.3--0.4 $\mu _{\text{B}}$ in LaFeAsO. \cite%
{CHLL+08,x:GABB+08}

Electron doping of the FeAs layers in LaFeAsO$_{1-x}$F$_{x}$ suppresses the
structural and magnetic transitions in favor of superconductivity already at 
$x$=0.03. \cite{DZXL+08} Also hole doping in Ba$_{1-2y}$K$_{2y}$Fe$_{2}$As$_{2}$
suppresses the structural and magnetic transition, \cite{RTJ08} but
this requires hole doping in excess of $y\approx $0.15.\cite{CRQB+08} Since
the superconducting transition occurs already for $y\approx $0.10, the
superconducting and striped AFM phases seem to coexist over a fairly wide
range of hole doping. \cite{CRQB+08,x:GABB+08} Although it is not clear
whether the superconductivity is mediated by AFM fluctuations or it competes
with magnetism, understanding merely the magnetic interactions is currently
of utmost importance.

A large number of electronic band-structure calculations using the local
spin-density approximation (LSDA) or generalized-gradient approximation now
exist for both 
LaFeAsO and BaFe$_{2}$As$_{2}$. \cite%
{DZXL+08,ITH08,YLHN+08,Yil08,MSJD08,SD08,Singh08_Ba,BDG08,OKZG+09,NPS08_comp}
The results obtained for LaFeAsO were reviewed and analyzed in
Ref.~\onlinecite{MJBK+08}. Although the Fe magnetic moment and the
stabilization energies of different magnetic solutions depend strongly on the
computational method and the exchange-correlation functional, as well as on
whether the experimental or calculated structure is used, all calculations
predict that stripe AFM order is the magnetic ground state in both parent
compounds. However, the calculated sublattice magnetizations are significantly
larger than the ones deduced from the neutron diffraction, $\mu$SR, and
M\"{o}ssbauer experiments. \cite{RTJS+08,CHLL+08,HQBG+08,x:GABB+08}

For both parent compounds, many authors (see, e.g., Refs.~%
\onlinecite{MSJD08,DZXL+08,OKZG+09}) have noticed a strong Fermi surface
(FS) nesting for the $\mathbf{q}$ vector which corresponds to stripe AFM
order between the Fe $d_{xz/yz}$-like hole sheet and one of the two electron
sheets. This nesting causes peaks in both the imaginary and real parts of
the non-interacting spin susceptibility, $\chi _{0}(\mathbf{q})$, at the
stripe $\mathbf{q.}$ Although electron doping of LaFeAsO suppresses the peak
and shifts it to an incommensurate wave vector,\cite{DZXL+08} it is widely
believed that stripe AFM order remains the LSDA ground state of LaFeAsO$%
_{1-x}$F$_{x}$. That merely 3\% electron doping suffices to destroy the
static stripe order, has been related to filling of the three-dimensional 3D
Fe $d_{3z^{2}-1}$-like band.\cite{MJBK+08} However, calculations which use the
experimental 
structure --such as those presented below-- place the top of the 
$d_{3z^{2}-1}$-like band several hundred meV below the Fermi level.

In this paper, we shall present LSDA calculations of moments and energies of
spin spirals in LaFeAsO$_{1-x}$F$_{x}$ and Ba$_{1-2y}$K$_{2y}$Fe$_{2}$As$_{2}$
 as functions of wave vector, $\mathbf{q}$, and doping, $x$ or $y$. We
find that upon increasing doping, stripe AFM order becomes unstable in favor
of an incommensurate spin-density wave (SDW).

Before getting to the spin spirals, we shall explain our computational
method and compare with previous results --and also present 
results involving spin-orbit coupling-- for the paramagnetic band
structures and the commensurate stripe and checkerboard SDWs for the parent
compounds.

\begin{figure}[tbp]
\begin{center}
\includegraphics[width=0.47\textwidth]{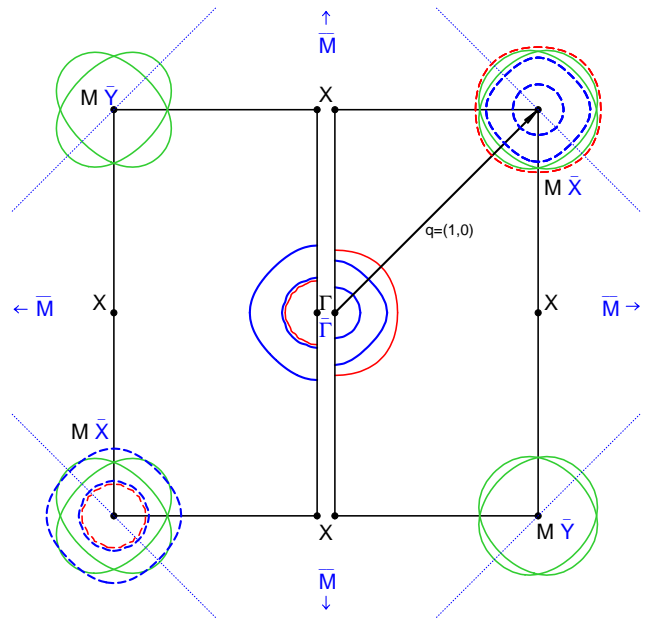}
\end{center}
\caption{(Color online) FS cross sections ($k_{z}\mathrm{=}0)$ for LaFeAsO
in the tetragonal Brillouin zone (BZ). The $a$ direction is horizontal, the
$b$ direction is vertical, and the coordinates of the (black) $\Gamma $, X, and
M points are as given in Fig.~\protect\ref{fig:bands}. The FS on the left-
and right-hand sides were calculated, respectively, with and without a 150 meV
downwards adjustment of $\protect\epsilon _{\protect\nu \,xy}$ in the LMTO
method. The $\Gamma $-centered $d_{xy}$-like hole cylinder is shown in thin
(red) line and the concentric $d_{yz,zx}$-like hole cylinders are shown in
thick (blue) lines. The two M-centered, electron cylinders shown in thin
(green) lines have super-ellipsoidal cross sections with main axes directed
toward $\Gamma$. The fact that a primitive translation of the square Fe
sublattice followed by mirroring in the Fe plane generates an Abelian
subgroup of the space group allows one to fold the band structure out to a
large one-formula-unit BZ. This moves the $d_{xz/yz}$-like hole cylinders
to the nearest-neighbor $\Gamma $ points, which are the (blue) \={M} corners
of the large BZ and have the coordinates $(1,1)\protect\pi \protect\sqrt{2}%
/a $ in the $\left( x,y\right) $-system. This also separates the electron
cylinders onto different M points, now called \={X} (1,0) and \={Y} (0,1)
(blue), such that the super-ellipses have their short axis pointing toward
the $\bar{\Gamma}$-centered $d_{xy}$-like hole cylinders and the long axis
pointing toward the \={M}-centered $d_{xz/yz}$-like hole cylinders.
Introducing a commensurate SDW with $\mathbf{q}=\left( 1,0\right) $ in the 
$\left( x,y\right)$ system, i.e., FM-ordered stripes in the $y$-direction and
AFM order in the $x$-direction, will shift the hole cylinders by $\mathbf{q}$.
This is shown by dashed lines around the bottom-left and top-right
corners, M=\={X}. The selection rules following from the glide mirror allow
coupling only between the $d_{xy}$ holes and the electron sheet with the
short axis along $\mathbf{q}$, and between the $d_{xz/yz}$ holes and the
electron sheet with the long axis along $\mathbf{q}$. Introducing a SDW with 
$\mathbf{q}=\left( 1,1\right)$, i.e., checkerboard AFM, folds the large BZ
back into the small tetragonal one. Here the nesting is between the $d_{xy}$%
-like and one of the $d_{xz/yz}$-like hole sheets and between two different
electron sheets. Not included in our spin-spiral calculations is the
spin-orbit coupling $\left( \protect\xi \sim 60\ \mathrm{meV}\right)$,
which invalidates the glide mirror and violates the selection rules.}
\label{fig:La_fs}
\end{figure}

\section{Computational method}

Our LSDA scalar-relativistic calculations for co-planar spin spirals in
LaFeAsO$_{1-x}$F$_{x}$ with $x$=0, 0.1, 0.2, and 0.3 and Ba$_{1-2y}$K$_{2y}$%
Fe$_{2}$As$_{2}$ with $y$=0, 0.1, 0.2, and 0.3 were carried out using the
linear muffin-tin orbital (LMTO) method in the atomic-sphere approximation
and including the combined correction term.\cite{And75} For the
exchange-correlation potential, we used the Perdew-Wang parametrization. 
\cite{PW92} Charge- and spin-self-consistent calculations were carried out
for all doping levels using the virtual-crystal approximation with
fractional atomic number of O or Ba. All our calculations were for the \emph{%
experimental} room-temperature crystal structures of the undoped compounds.%
\cite{KWHH08,RTJS+08} This is important. Whereas the primitive cell of
LaFeAsO holds two formula units and is tetragonal ($P4/nmm$), that of BaFe$%
_{2}$As$_{2}$ holds one unit and is body-centered tetragonal (bct) ($I4/mmm$).
The lattice constants as well as the positions and radii of the
space-filling atomic and empty (E) LMTO spheres are given in Table \ref%
{TableStructure}.

\begin{table}[tbp]
\caption{Lattice and LMTO spheres}
\label{TableStructure}
\begin{ruledtabular}
\begin{tabular}{ccc}
Sphere & Wycoff position & Radius (\AA)\\ 
\hline
\multicolumn{3}{c}{LaFeAsO ($P4/nmm$), $a=b=4.04$ \AA, $c=8.74$ \AA} \\ 
Fe & $2b$  (0.25, 0.75, 0.50) & 1.41 \\ 
As & $2c$  (0.25, 0.25, 0.651) & 1.48 \\ 
La & $2c$  (0.25, 0.25, 0.142) & 1.71 \\ 
O & $2a$  (0.25, 0.75, 0) & 1.13 \\ 
E$_{1}$ & $2c$  (0.25, 0.25, 0.407) & 1.08 \\ 
E$_{2}$ & $4f$  (0.25, 0.75, 0.266) & 1.04 \\ 
E$_{3}$ & $2c$  (0.25, 0.25, $-$.115)  & 0.97 \\ 
\hline
\multicolumn{3}{c}{BaFe$_2$As$_2$ ($I4/mmm$), $a=b=3.96$ \AA, $c=13.02$ \AA} \\ 
Fe & $4d$  (0, 0.5, 0.25) & 1.38 \\ 
As & $4e$  (0, 0, 0.355) & 1.46 \\ 
Ba & $2a$  (0, 0, 0) & 2.01 \\ 
E$_{1}$ & $4e$  (0, 0, 0.195) & 0.99 \\ 
E$_{2}$ & $2b$  (0, 0, 0.50) & 0.78 \\ 
E$_{3}$ & $8g$  (0.5, 0, $-$.105) & 0.84 \\
\end{tabular}
\end{ruledtabular}
\end{table}

\begin{figure*}[tbp]
\begin{center}
\includegraphics[width=0.98\textwidth]{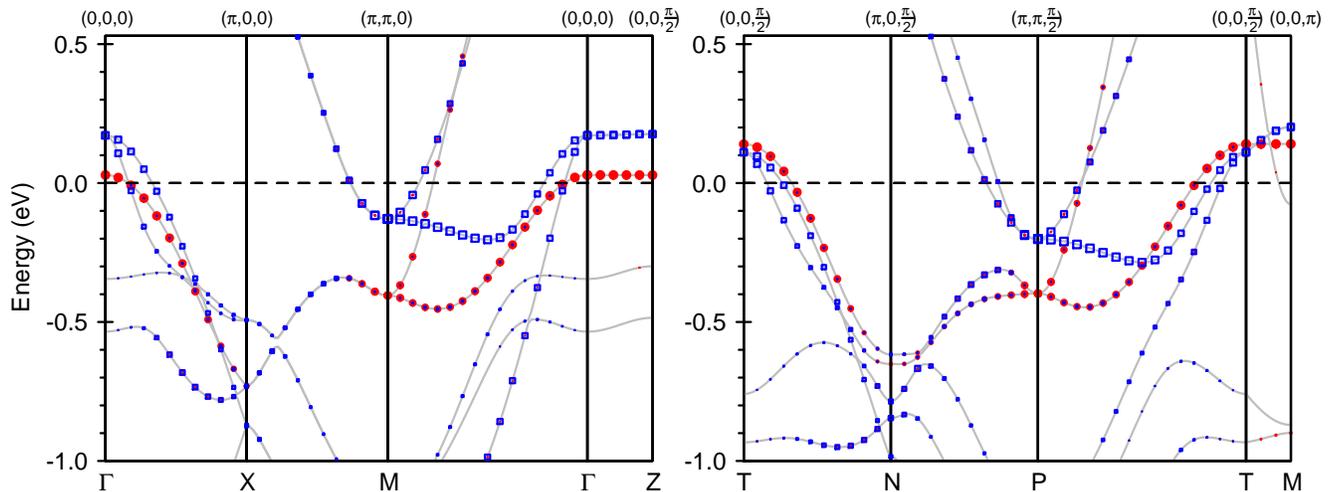}
\end{center}
\caption{(Color online) Band structures of undoped LaFeAsO (left panel) and
BaFe$_{2}$As$_{2}$ (right panel) obtained from non-spin-polarized
calculations for the experimental structure. The size of (red) circles and
(blue) squares is proportional to, respectively, the Fe $d_{xy}$ and
$d_{yz,zx}$ partial-wave character. The coordinates shown at the bottom are 
in the tetragonal $\left( a,b,c\right) $-system where the cell contains two
Fe atoms, i.e., in units of $\left( 1/a,1/b\mathrm{=}1/a,1/c\right)$. The
Fermi level is the zero of energy.}
\label{fig:bands}
\end{figure*}

For the Fe $d$ orbitals, such as $d_{xy}$, we use the $x$ and $y$ axes (not
to be confused with the levels of electron and hole doping) which span the
quadratic Fe sublattice and are therefore turned 45${}^\circ$
 with respect to the tetragonal $a$ and $b$ axes. The lobes of the Fe 
$d_{xy}$ orbital thus point toward the projections of the As sublattice onto
the Fe plane, while $d_{x^{2}-y^{2}}$ points toward the nearest Fe
neighbors. Or in other words: the Fe-Fe $dd\pi $ interaction involves the 
$d_{xy}$ orbitals and the Fe-Fe $dd\sigma $ interaction involves the
$d_{x^{2}-y^{2}}$ 
orbitals. This convention is the same as the one used for the Cu orbitals in
the high-temperature superconducting cuprates. The distance, $a/\sqrt{2}$,
between Fe nearest neighbors is 2.85 \AA\ in LaFeAsO and 2.80 \AA\ in
BaFe$_{2}$As$_{2}$.

Our band structures and Fermi surfaces for undoped paramagnetic LaFeAsO\
and BaFe$_{2}$As$_{2}$ obtained from spin-restricted LDA LMTO calculations
agree well with those obtained with the same method and presented in Ref.~%
\onlinecite{NPS08_comp}.

However, for LaFeAsO the LMTO Fermi surface differs from that obtained with
the full-potential linear augmented plane-wave (LAPW) method:%
\cite{MJBK+08} Whereas LMTO finds the two innermost $\Gamma $-centered
hole cylinders to be $d_{xz,yz}$ like and the outermost $d_{xy}$ like, LAPW
finds the opposite order. Since LAPW is computationally more cumbersome and
accurate than LMTO, we used LMTO but applied an external crystal field which
shifted the energy $\left( \epsilon _{\nu \,xy}\right) $ of the Fe $d_{xy}$
partial wave downwards by 150 meV. This adjustment brought the LMTO and LAPW
band structures into almost complete agreement. The adjusted and unadjusted
Fermi surfaces are shown respectively on the left and right-hand sides of
Fig.~\ref{fig:La_fs}. In the following, all results presented for LaFeAsO$%
_{1-x}$F$_{x}$ --such as the paramagnetic bands on the left-hand side of
Fig.~\ref{fig:bands} and the spin-spiral moments and energies in
Fig.~\ref{fig:emq_LaBa}-- are those obtained with LMTO and the $d_{xy}$-energy 
downshifted, unless otherwise stated.

For BaFe$_{2}$As$_{2}$, the LMTO LDA band structure shown on the right-hand
side of Fig.~\ref{fig:bands} agrees very well with the LAPW one. For that
reason, our LMTO calculations for Ba$_{1-2y}$K$_{2y}$Fe$_{2}$As$_{2}$ were
all performed without any adjustment.

Finally, we calculated effects of spin-orbit coupling (Fermi surface
splittings and magnetocrystalline energies) using the spin-polarized
relativistic LMTO method \cite{APSY95} and the experimental structure.

\section{\label{sec:lsda}Paramagnetic energy bands}

The paramagnetic scalar-relativistic bands for undoped LaFeAsO are shown
near the Fermi level at the left-hand side of Fig.~\ref{fig:bands}. We see
that at $\Gamma$, the top of the $d_{xz/yz}$ band (blue squares) is $\sim $%
180 meV while that of the $d_{xy}$ band (red spheres) is merely $\sim $30
meV above $\varepsilon _{F}$. Without the 150 meV shift, these levels would
have been nearly degenerate, and since the mass of the $d_{xy}$ band is
higher that those of the two $d_{xz/yz}$ bands, the cylindrical $d_{xy}$
sheet would have been the widest, as shown in the right-hand side of
Fig.~\ref{fig:La_fs}. Although spin-orbit coupling cannot be included in 
spin-spiral calculations, we mention that it splits the degenerate top of
the $d_{xz/yz}$-like band by 50 meV, a value consistent with $\xi _{\mathrm{%
Fe}\,3d}\approx 60$ meV and a 20\% reduction due to by-mixing of As $p$
character.

With an even number of electrons (Fe $d^{6}$), the sum of the volumes of the
three hole sheets equals that of the electron sheets. Of those, there are
two equivalent cylinders centered at M and turned 90${}^\circ$
 with respect to each other (thin green). Their cross sections are
superellipsoidal with main axes pointing toward $\Gamma $. The
superellipsoidal cross section arises because these sheets result from a 
$d_{xz}$ or $d_{yz}$ band hybridizing with a lower-lying $d_{xy}$-like band.
\cite{OKA&Lilia} The main character is $d_{xz}$ or $d_{yz}$ near the short
axis and $d_{xy}$ near the long axis. These electron cylinders have more As
$p$ character and more $k_{z}$-dispersion than the hole cylinders, and are 
therefore more warped.

The fact that a primitive translation of the square Fe sublattice followed
by mirroring in the Fe plane generates an Abelian subgroup of the $P4/nmm$
space group allows one to fold the band structure out to the large Brillouin
zone (BZ) well known from the cuprates. \cite{Lee,OKA&Lilia}
This folding out has the advantage of separating the $d_{xy}$ cylinders
from the $d_{xz/yz}$ ones by placing them at respectively $\bar{\Gamma}$
(0,0) and \={M} (1,1) in the $\left( x,y\right)$ system. It also separates
the two electron pockets from each other by placing them at, respectively,
\={X} (1,0) and \={Y} (0,1) with the long axis pointing toward \={M}.

Note that in order to distinguish the tetragonal $\left( a,b,c\right)$%
-directed coordinates in reciprocal space from the quadratic $\left(
x,y,z\parallel c\right) $-directed ones, we use the units $\left(
1/a,1/a,1/c\right) $ in the former case and $\left( \sqrt{2}\pi /a,\sqrt{2}%
\pi /a,\pi /c\right) $ in the latter case. Hence, tetragonal (quadratic)
reciprocal-space coordinates are recognized by the presence (absence) of the
factor $\pi$.

Spin-orbit coupling does not commute with the above-mentioned glide mirror
and will therefore split the crossing between the $d_{xy}$ and one of the $%
d_{xz/yz}$ hole bands by about $50$ meV, as well as the crossings between
two electron bands as has been observed in LaFePO. \cite{CFCA+08}

At the right-hand side of Fig.~\ref{fig:bands}, we show the paramagnetic
bands near the Fermi level for undoped BaFe$_{2}$As$_{2}$. Since the As-Fe$%
_{2}$-As layers are separated by a thin Ba layer, rather than by a thick La-O%
$_{2}$-La layer, and the AS atoms along the z axis are stacked on top of each
other the band structure of BaFe$_{2}$As$_{2}$ disperses more in
the $z$ direction than that of LaFeAsO. But apart from that, the band
structures are very similar. For ease of comparison with the LaFeAsO band
structure at the left-hand side of the figure, we have chosen the same route
in $\left( k_{x},k_{y}\right) $ space, but have taken $k_{z}=\pi /\left(
2c\right) $ --except in the very last panel-- because this choice makes the
amplitude of the Bloch waves vanish in the Ba-plane and thus minimizes the
effects of $k_{z}$ dispersion. Along $\left( 0,0,k_{z}\right)$, the $d_{xy}$%
-like band is 140 meV above $\varepsilon _{F}$ and does not disperse with $%
k_{z}$, whereas the doubly degenerate $d_{xz/yz}$-like band disperses from
40 to 200 meV above $\varepsilon _{F}$ with $k_{z}$ increasing from 0 to $%
\pi /c$, and thus goes from below to above the top of the $d_{xy}$ band. The
hole sheets thus remain cylinders although the $d_{xz/yz}$-like sheets are
significantly warped. Near $\left( 0,0,\pi /c\right)$, we see a band with
strong $k_{z}$-dispersion dip below $\varepsilon _{F}$. This band is As
$p_{z}$ like and cannot hybridize with the $d_{xy}$-like band, but only with 
the $d_{xz/yz}$-like band, but not along $\left( 0,0,k_{z}\right) $. This
hybridization gives rise to an intricate shape of the $d_{xz/yz}$ bands near 
$\varepsilon _{F}$ for $k_{z}\neq \pi /\left( 2c\right) $ and is discussed
in Ref.~\onlinecite{OKA&Lilia}. A further difference with the LaFeAsO bands is
that the electron cylinders around $\left( \pi ,\pi ,k_{z}\right) $ are not
degenerate along the $a$ and $b$ directions and that the $k_{z}$ dispersion
of the $d_{xy}$-like component is as large as 150 meV due to by-mixing of As 
$p_{z}$ character.

\begin{figure}[tbp]
\begin{center}
\includegraphics[width=0.48\textwidth]{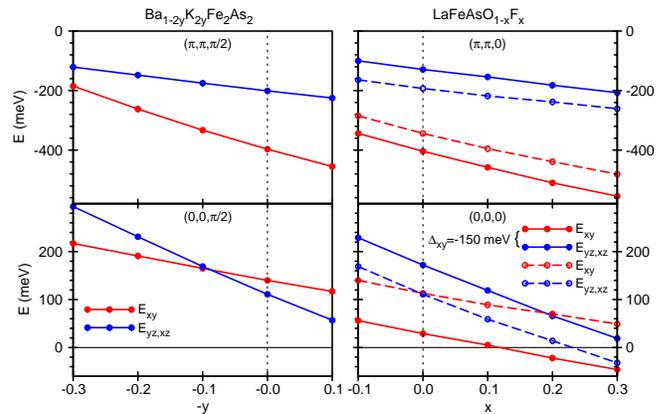}
\end{center}
\caption{(Color online) Band positions as functions of doping in the
range $-30\%\leq x=-y\leq 30\%$ for Ba$_{1-2y}$K$_{2y}$Fe$_{2}$As$_{2}$
(left) and LaFeAsO$_{1-x}$F$_{x}$ (right).}
\label{fig:doping}
\end{figure}

The main effect of electron doping on the band structure is to move the
Fermi level up or --equivalently-- to move the bands down with respect to
the Fermi level. This is clearly seen in Fig.~\ref{fig:doping} where we show
dependence on doping -- ranging from $-30\%$ in Ba$_{1-2y}$K$_{2y}$Fe$_{2}$%
As$_{2}$ to $+30\%$ in LaFeAsO$_{1-x}$F$_{x}$ -- of the top of the $d_{xy}$%
-like (red points) and $d_{xz/yz}$-like (blue squares) hole bands at (0, 0,
$\pi/2$, or 0) and the bottoms of the $d_{xy}$- and 
$d_{xz}$- or $d_{yz}$-like electron bands at ($\pi$ ,$\pi$ ,$\pi/2$, or 0). We
note, first of all, that the dominating rigid
shift of these levels is roughly continuous when passing from hole doping in
Ba$_{1-2y}$K$_{2y}$Fe$_{2}$As$_{2}$ to electron doping in LaFeAsO$_{1-x}$F$%
_{x}$ (and even more so had we not corrected $\epsilon _{\nu \,xy}$ for the
latter). Second, we note that 11\% electron doping fills the $d_{xy}$
band, and 33\% fills the $d_{xz/yz}$ band, and that no other Fe $d$-like band
gets 
filled or emptied in the $\pm 30\%$ doping range. Within this range, the
bands move by about 300 meV, corresponding to an average density of states
of order 2 electrons/Fe/eV. The deviation from rigid-band behavior, i.e.,
parallel movement of all levels, is roughly 50\% of this.

\section{\label{sec:commensurate}Magnetic moments and energies of collinear, commensurate SDWs in the parent compounds}

For LaFeAsO, our LMTO calculations with imposed FM order converge to a
nonmagnetic solution. For stripe and checkerboard AFM orders we calculate
magnetic stabilization energies of, respectively, $78$ and $39$ meV/Fe, values
which compare well with those, $84$ and $26$ meV/Fe, obtained from LAPW
calculations for the experimental structure.\cite{MJBK+08} Our
sublattice magnetizations (Fe moments) of 1.3 for stripe and 1.2 $\mu _{%
\text{B}}$/Fe for checkerboard AFM order are somewhat smaller than those, 
1.8 and 1.5 $\mu _{\text{B}}$/Fe, found with LAPW.\cite{MJBK+08} This
could be due to integrating the spin-density over different regions, i.e.,
LMTO and LAPW spheres have different sizes. The LSDA band structure and
Fermi surface for stripe order are in a good agreement with those obtained
with LAPW.\cite{YLHN+08} Finally, we mention that without the
downwards $\epsilon _{\nu \,xy}$ shift, LMTO yields slightly larger
energies, 95 and 53 meV/Fe, and moments, 1.4 and 1.4 $\mu _{\text{B}}$%
/Fe.

For BaFe$_{2}$As$_{2}$, we also find no FM solution. For stripe order, we
find a somewhat lower stabilization energy, 62 meV/Fe, than for LaFeAsO
but the same moment, 1.3 $\mu _{\text{B}}$/Fe. Using LAPW
(Ref~\onlinecite{WIEN2K}) 
instead of LMTO, we find the same stabilization energy, 85 meV/Fe, as for
LaFeAsO, and a slightly smaller moment, 1.7 $\mu _{\mathrm{B}}$/Fe. The
calculated LMTO and LAPW band structures for the striped phase agree
reasonably well. LMTO calculations performed for the experimental
low-temperature orthorhombic structure\cite{RTJS+08} with the FM stripes
oriented either along the longer or the shorter axis result in a lower
total energy for the latter. This is in accord with the experimental data in
Ref.~\onlinecite{HQBG+08} and a previous calculation.\cite{ITH08}

For stripes in both LaFeAsO and BaFe$_{2}$As$_{2}$, the maximum exchange
splitting, i.e., that of degenerate bands, $\varepsilon _{j\,\mathbf{k}}$ and 
$\varepsilon _{j^{\prime }\,\mathbf{k+q}}$, with the same dominant $d$
character,\cite{OKA&Lilia} is $\Delta \approx 1.2$ eV, and this is
consistent with an Fe moment of $M\approx 1.3$ $\mu _{\mathrm{B}}$/Fe and
Stoner theory: $\Delta =MI$, with the usual value of the Fe effective Stoner
parameter ($\sim $ Hunds rule $J_{H})$: $I\approx 0.9$ eV. \cite{Gun76,PKA76}
Since $\Delta$ 
is an order of magnitude larger than the energies of the (paramagnetic,
un-coupled) electron and hole pockets with respect to the Fermi energy, but
considerably smaller that the sub-band widths --which are several eV-- the
strength of the LSDA exchange coupling is \emph{intermediate} when the
magnetic order is stripe or checkerboard. This means that neither
Fermi surface nesting, nor the difference between the band structures of
pure LaFeAsO and pure BaFe$_{2}$As$_{2}$, nor the 150 meV adjustment for
LaFeAsO, strongly affects the magnetic moments and energies.

Although the spin-orbit coupling in the Fe $3d$ shell is weak ($\xi _{%
  \mathrm{Fe}\,3d}\approx 60$ meV), it does lead to a distinct
magnetocrystalline anisotropy with the Fe moment lying in the plane and
perpendicular to the FM stripes. This has been shown by neutron diffraction
for BaFe$_{2}$As$_{2}$.\cite{HQBG+08} Our relativistic LMTO calculations agree
with this: For LaFeAsO we find that it costs 0.27 meV/Fe to turn the moment in
the plane from the easy to the FM-stripe direction, and an additional 0.13
meV/Fe to turn the moment perpendicular to the plane. For BaFe$_{2}$As$_{2}$,
the corresponding energies are 0.16 and 0.04 meV/Fe, i.e., the anisotropies
are smaller. 

\begin{figure*}[tbp]
\begin{center}
\includegraphics[width=0.96\textwidth]{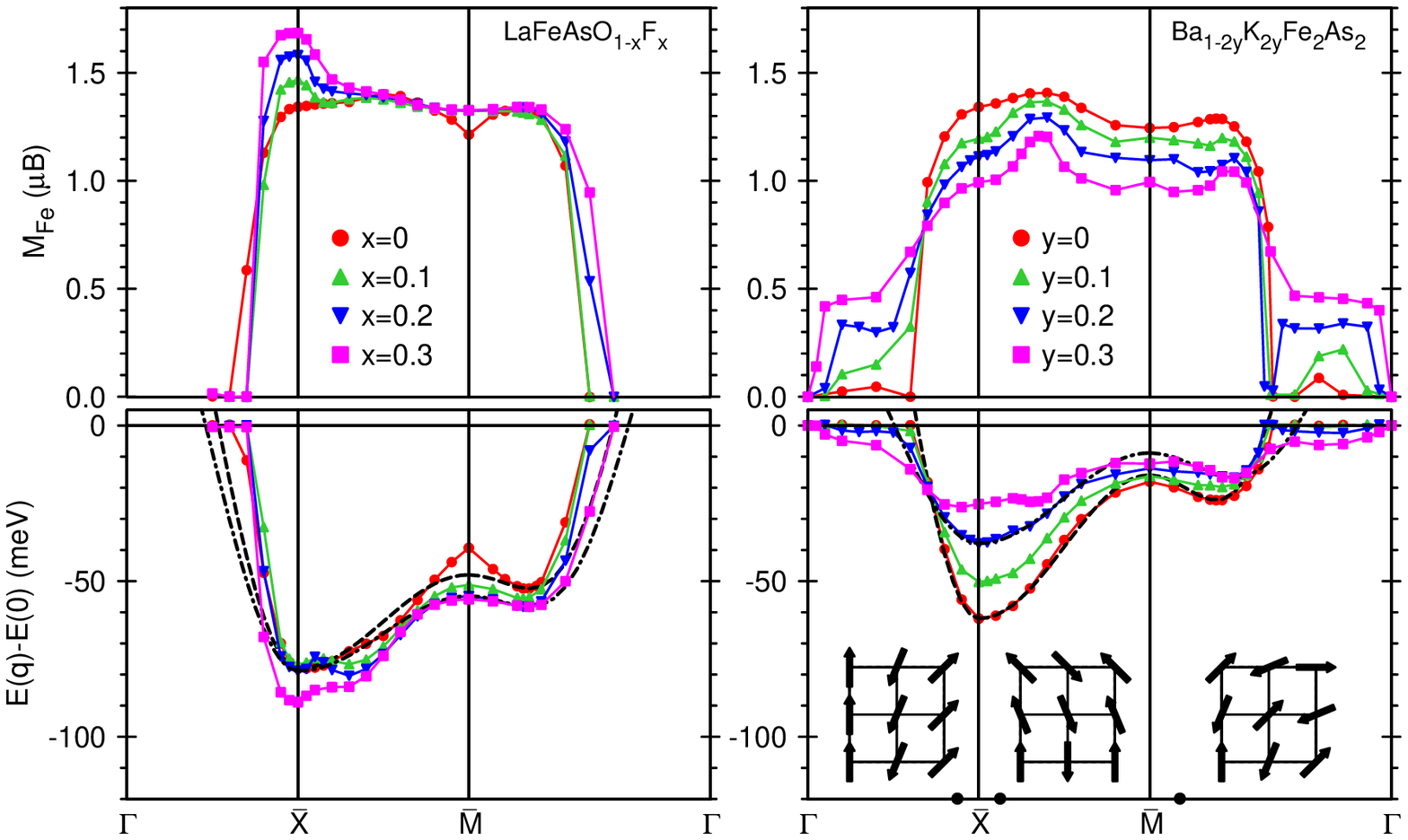}
\end{center}
\caption{(Color online) Magnetic moments (upper panels) and energies (lower
panels) per Fe of in-plane spin spirals as functions of $\mathbf{q=(}%
q_{x},q_{y},0)$ along the boundaries $\bar{\Gamma}\left( 0,0\right) $ $-$ $%
\bar{\mathrm{X}}\left( 1,0\right) $ $-$ $\bar{\mathrm{M}\left( 1,1\right) 
\text{ }}-$ $\bar{\Gamma}\left( 0,0\right) $ of the irreducible part of the
large BZ for different electron $\left( x\right) $ and hole $\left(
y\right) $ dopings. The energies of the $j_{1}$--$j_{2}$ Heisenberg model
for $x$($y$)=0 and 0.2 are given by, respectively, dashed and dash-dotted
lines. Representative real-space spin structures are shown at the bottom
right for the $\mathbf{q}$ vectors denoted by dots.}
\label{fig:emq_LaBa}
\end{figure*}

\section{\label{sec:ssc}Magnetic moments and energies of spin spirals in the doped compounds}

\subsection{Spin spirals}

A spin spiral is characterized by the following properties: Upon a lattice
translation $\mathbf{t}$, the magnitude of the magnetization density and its
projection onto a global $z$ direction are unchanged, but the projection
onto the perpendicular $\left(x,y\right)$ plane rotates by an angle $\varphi
\left( \mathbf{t}\right) $ proportional to the projection of the 
translation onto the wave vector $\mathbf{q}$ of the spin spiral. The
magnetization-density of a spin spiral thus satisfies the equation 
\begin{eqnarray*}
&&\mathbf{M}\left( \mathbf{r+t,q}\right) =\mathbf{M}\left( \mathbf{r,q}%
\right) \\
&&\quad \cdot \left\{ \mathbf{\hat{z}\hat{z}}\cos \theta +\left[ \mathbf{%
\hat{x}\hat{x}}\cos \varphi \left( \mathbf{t}\right) +\mathbf{\hat{y}\hat{y}}%
\sin \varphi \left( \mathbf{t}\right) \right] \sin \theta \right\} \\
&&\mathrm{with}\;\;\varphi \left( \mathbf{t}\right) =\pi \mathbf{q\cdot t}.
\end{eqnarray*}%
(The factor $\pi $ merely follows from the convention chosen in
Sec.~\ref{sec:lsda} for $\mathbf{t}$ referring to the Fe
sublattice.) Examples of  
such spin spirals are shown by the solid arrows in Fig.~\ref{fig:emq_LaBa}
at the bottom of the right-hand side. Note that for Fe atoms lying along
rows perpendicular to $\mathbf{q}$, the alignment is FM.

In order to solve the one-electron problem in the presence of such a spin
spiral, one may span the one-electron Hilbert space by a complete set of
localized orbitals, $\phi _{j}\left( \mathbf{r}-\mathbf{t}\right)$, times
pure spin-functions, $\chi _{\mathbf{t}}\left( \sigma \right) =\alpha _{%
\mathbf{t}}\left( \sigma \right) $ or $\beta _{\mathbf{t}}\left( \sigma
\right)$, whose quantization direction is chosen along the \emph{local}
direction of the magnetization. In this representation, the one-electron
Hamiltonian \emph{without} spin-orbit coupling is translationally invariant,
albeit with $\mathbf{q}$-dependent hopping integrals, so that there is \emph{%
no} coupling between Bloch sums, $\sum_{\mathbf{t}}\phi _{j}\left( \mathbf{r}%
-\mathbf{t}\right) \chi _{\mathbf{t}}\left( \sigma \right) \exp \left( \pi i%
\mathbf{k\cdot t}\right)$, with different Bloch vectors. As a consequence,
the one-electron problem can be solved for \emph{any} $\mathbf{q}$, \emph{%
without} increasing the size of the primitive cell, provided that spin-orbit
coupling is neglected.\cite{San91a} This, together with the LSDA, enables
simple calculation of spin-spiral moments and total energies.

In our LMTO calculations, the localized orbitals were taken to be the
partial waves truncated outside their sphere. This means that we forced the
direction of magnetization to be constant inside each sphere. The moment
that we quote is the one integrated over an Fe sphere. We considered spin
spirals for which the magnetization is \emph{in} the Fe $\left( x,y\right)$
plane, i.e., $\theta =\pi /2$ and first took $\mathbf{q}$ to lie \emph{in}
the plane and $\mathbf{t}$ to span the square Fe sublattice.
To achieve this in calculations employing the tetragonal translational cell,
the phases $\varphi_{i}$, which determine the magnetization directions in two
Fe spheres at positions $\bm{\tau}_i$, were fixed to
$\varphi_{i}=\mathbf{q}\cdot\bm{\tau}_i$.

The spin spiral with $\mathbf{q}=\mathrm{\bar{\Gamma}}(0,0)$ produces FM
order. In the spin spiral with $\mathbf{q}=\mathrm{\bar{X}}(1,0)$, the
moments rotates by $\pi $ upon translation by $\mathbf{t=}\left( 1,0\right)
, $ and by 0 upon translation by $\mathbf{t=}\left( 0,1\right)$, i.e., the
order is stripe with AFM alignment of nearest-neighbor moments along the $x$
direction and FM alignment along the $y$ direction. In the spin spiral with 
$\mathbf{q}=\mathrm{\bar{M}}\left( 1,1\right)$, the moments rotate by $\pi $
upon translation by $\mathbf{t}=\left( 1,0\right)$, as well as by $\mathbf{t%
}=\left( 0,1\right)$, i.e., the order is checkerboard with all four
nearest-neighbor moments antiparallel and all four second-nearest moments
parallel. These spin spirals with $\mathbf{q}$ at high-symmetry points are
all collinear and commensurate.

Noncollinear and incommensurate spin spirals with $\mathbf{q}$ near --but
not at-- \={X} and \={M} are illustrated at the bottom of the right-hand
side of Fig.~\ref{fig:emq_LaBa}. When $\mathbf{q}$ is on the $\bar{\Gamma}%
\mathrm{\bar{X}}$ line, the order along the $y$ direction remains FM, while
going from one Fe to the next in the positive $x$ direction, the moment
rotates by between $0$ and $\pi$. When $\mathbf{q}$ is on the $\mathrm{\bar{%
X}\bar{M}}$ line, the order along the $x$ direction is AFM, while upon going
from one Fe to the next in the positive $y$ direction, the moment rotates by
between $0$ and $\pi$. When finally $\mathbf{q}$ is on the $\mathrm{\bar{M}}%
\bar{\Gamma}$ line, the moment rotates by the same angle, lying between $0$
and $\pi$, regardless of whether $\mathbf{t}=\left( 1,0\right) $ or $\left(
0,1\right)$, and the order along the second-nearest-neighbor direction
perpendicular to $\mathbf{q}$, i.e., for $\mathbf{t}=\left( 1,-1\right)$, is
FM.

We now discuss the calculated results shown in this figure for the magnetic
moments (top) and energies (bottom) of spin spirals in electron-(left) and
hole-doped (right) compounds as functions of $\mathbf{q}$ along the
triangular boundary $\bar{\Gamma}$-$\bar{\mathrm{X}}$-$\mathrm{\bar{M}}$-$%
\bar{\Gamma}$ of the irreducible part of the large BZ (see
Fig.~\ref{fig:La_fs}). 

\subsection{\label{sec:sla}Pure and electron-doped LaFeAsO$_{1-x}$F$_{x}$}

For undoped LaFeAsO, i.e. for $x=0$ (red dots), the lowest energy is reached
at the $\bar{\mathrm{X}}$ point, i.e., for stripe AFM order. This agrees
with the results of previous calculations \cite{Yil08,ITH08,OKZG+09} and
experimental data.\cite{CHLL+08} When moving from \={X}$\left( 1,0\right) $
toward $\bar{\Gamma}\left( 0,0\right)$, the AFM order \emph{between}
nearest-neighbor Fe rows along $y$ is destroyed, and this leads to a rapid
decrease in moment and increase in energy; for $\left\vert \mathbf{q}%
\right\vert \equiv q<0.6$ the self-consistent solution is nonmagnetic. When
moving from \={X}$\left( 1,0\right) $ toward \={M}$\left( 1,1\right) $, the
FM order \emph{in} the Fe rows along $y$ is destroyed and becomes AFM at 
$\bar{\mathrm{M}}$. Whereas the moment at first remains nearly constant at
1.4 $\mu _{\text{B}}$, but finally decreases to 1.2 $\mu _{\text{B}}$ at $%
\bar{\mathrm{M}}$, the energy increases nearly parabolically from $-78$ to 
$-39$ meV. Moving from \={M}$\left( 1,1\right) $ toward $\bar{\Gamma}\left(
0,0\right)$, the nearest-neighbor AFM order develops toward FM order by
preserving the FM order between second-nearest neighbors along $\mathbf{t}%
=\left( 1,-1\right) $ but destroying the one along $\mathbf{t}=\left(
1,1\right)$. Hereby the moment first increases slightly, but then decreases
rapidly and vanishes when $q<0.7$. The energy falls to a local minimum at 
$q\approx 1$, and then increases rapidly for $q$ decreasing to 0.7 where the
moment disappears.

As shown by the dashed curve in Fig.~\ref{fig:emq_LaBa}, for $q\gtrsim
0.7 $ our calculated $E\left( \mathbf{q}\right) $ for undoped LaFeAsO is
approximated reasonably well by a classical Heisenberg model on the square
lattice with AFM exchange coupling constants $j_{1}$ and $j_{2}$ between,
respectively, nearest and next-nearest neighbors. We obtain the ratio 
$j_{2}/j_{1}=0.71~\left( >1/2\right) $ by fitting to the $\mathbf{q}$
position of the local minimum along \={M}$\bar{\Gamma}$ and then obtain the
values $j_{1}=81$ and $j_{2}=57$ meV by fitting the energy difference $%
E_{\min }\left( \mathrm{\bar{M}}\bar{\Gamma}\right) -E\left( \mathrm{\bar{X}}%
\right) $ and the value $S\equiv \frac{1}{2}M\left( \mathrm{\bar{X}}\right)
=0.67$. The $\mathbf{q}$-independent constant is finally chosen such that
the Heisenberg model fits the calculated energy at $\mathrm{\bar{X}}$ (and
the minimum along \={M}$\bar{\Gamma}$). Our values of $j_{1}$ and $j_{2}$
are comparable to those obtained from Fig.~3 in Ref.~\onlinecite{Yil08}
when interpolated to $M\left( \mathrm{\bar{X}}\right) =1.3$ $\mu _{\text{B}}$,
although our $j_{2}/j_{1}$ is somewhat higher. The anisotropic exchange
coupling constants calculated in Ref.~\onlinecite{YLHN+08} are meaningful
only for small deviations from stripe AFM order, that is for $\mathbf{q}$
near \={X}, and cannot be directly compared with our effective $j_{1}$ and 
$j_{2}$. Nevertheless, our $j_{1}S^{2}=420$ K and $j_{2}S^{2}=297$ K are of
the same order of magnitude as $-J_{1}^{\perp }=550$ K and $-J_{2}^{\perp
}=260$ K of Ref.~\onlinecite{YLHN+08}. This indicates that the three
different approaches result in the comparable exchange interactions.
Although the overall shape of the dashed $E\left( \mathbf{q}\right)$ curve
given by the Heisenberg model is similar to the red-dotted one obtained from
our LSDA calculation, discrepancies can be clearly seen even in the part of
the BZ where the Fe moment remains nearly constant: The calculated energy is
higher in the vicinity of the $\bar{\mathrm{M}}$ point and it increases far
more rapidly when going from \={X} toward $\bar{\Gamma}$.

Our calculations reveal that a new local minimum of $E(\mathbf{q})$ 
develops along \={X}\={M} near $(1,0.3)$ when the electron doping exceeds
about $5\%$. For $0.10<x<0.25$, this minimum is deeper than the one a \={X},
i.e. it is the global minimum, and this means that stripe order is unstable
against formation of an incommensurate noncollinear SDW. This is in line
with experimental phase diagrams for REFeAsO$_{1-x}$F$_{x}$ compounds which
show that magnetic order is rapidly suppressed by F doping.\cite{DZXL+08}
For $x=0.3$ we find that the energy minimum has returned to $\bar{\mathrm{X}}$,
although a well-defined shoulder can still be seen at $(1,0.3)$. This
instability toward an incommensurate SDW cannot be reproduced by fitting
to the $j_{1}$-$j_{2}$ Heisenberg model, as is clearly seen by comparison of
the dot-dashed and blue $\blacktriangledown$ curves.

Electron doping is seen to increase the moment for stripe AFM order from
1.4 to 1.7 $\mu _{\text{B}}$ for $x=0.3$. But this increase is localized to 
$\mathbf{q}$ being close to \={X}.

The destabilization of stripe AFM order with electron doping seems to be
caused by occupation of a narrow peak of the density of states (DOS), which
in undoped LaFeAsO is $\sim $150 meV above the Fermi level ($\varepsilon
_{F}$). The band responsible for this peak is the paramagnetic Fe $d_{yz}$ 
band which hardly disperses in the $k_{x}$ direction and stays within $\pm
200$ meV of the Fermi level over a region of $\mathbf{k}$ space near the
entire \={M}\={Y} line. This is the band seen in Fig.~\ref{fig:bands} to
connect the $d_{yz}$-like hole pocket at \={M} with the superellipsoidal
electron pocket at \={Y}. Now, FM stripes in the $y$ direction with AFM
order along $x$ will couple states at $\mathbf{k}$ with those at $\mathbf{%
k+q=k}+\left( 1,0\right)$, i.e. will fold the large BZ perpendicular to the 
$y$ direction, thus placing \={M} on top of \={Y}, and exchange-split states
with the same Fe $d$ character by $\sim \pm 0.5$ eV. As a result, the flat 
$d_{yz}$ band will have its upper minority-spin $d_{yz}$ band $\sim$150 meV
above $\varepsilon _{F}$ in the undoped compound.\cite{OKA&Lilia} Since the
DOS near $\varepsilon _{F}$ is very low in the stripe-ordered undoped
compound, not much electron doping is needed to occupy part of the flat 
$d_{yz}$ minority-spin band. For $\mathbf{q}$ moving away from \={X} in the
perpendicular direction, i.e., toward \={M}, the DOS peak soon splits in
two. The concomitant decrease in band energy for dopings such that 
$\varepsilon _{F}$ is in the valley between the subpeaks compensates for the
decrease in negative magnetic exchange energy $\left( -\frac{1}{4}%
M^{2}I\right) $ caused by the decrease in magnetization seen in the upper
left-hand side of Fig.~\ref{fig:emq_LaBa}. The magnetization decreases
because the nesting is less good (less phase space available for gapping)
when $\mathbf{q}$ moves so far away from \={X} in the perpendicular
direction that $\mathbf{q}$ does not so well translate the $\mathbf{k}$ tube
around \={M}\={Y} in which the $d_{yz}$ band is flat onto itself. The
energy minimum finally shifts back to $\mathbf{q}=\bar{\mathrm{X}\left(
1,0\right) }$ once the electron doping is so high $\left( >25\%\right) $
that the narrow DOS peak is completely filled, i.e., when $\varepsilon _{F}$
is above the flat part of the minority-spin $d_{yz}$ band.

Recently, a commensurate-to-incommensurate transition with increasing $x$ in
LaFeAsO$_{1-x}$F$_{x}$ has been reported in Ref.~\onlinecite{x:SDSB+08} on
the base of full-potential LAPW spin-spiral calculations. These calculations
were however performed using \emph{calculated} As positions and can
therefore not be compared directly our calculations based on the
experimental structure.

Returning now to the undoped compound and $\mathbf{q}$ moving away from \={X}
toward \={M}, $\varepsilon _{F}$ is below both subpeaks and the
Heisenberg-type dependence of $E(\mathbf{q})$ persists until in the
vicinity of $\bar{\mathrm{M}}$ a huge DOS peak appears just below 
$\varepsilon _{F}$.

For stripe-ordered pure LaFeAsO, another narrow DOS peak exists 270 meV
above $\varepsilon _{F}$ and arises from the upper minority-spin $d_{xy}$%
-like band being split from the lower minority-spin band due to folding of
the \={M}-centered hole pocket onto the \={Y}-centered electron pocket.\cite%
{OKA&Lilia}

We estimated the strength of the \emph{interlayer} exchange coupling by
performing calculations for spin spirals with $\mathbf{q}$ not lying in the
Fe plane. Specifically we took $\mathbf{q}=(1,0,q_{z})$ corresponding to
stripe order in each layer and rotation of the magnetic moments between
layers. These calculations resulted in a very small change in energy when
changing the alignment of the Fe moments along the $c$ direction from
ferromagnetic 
to antiferromagnetic. A weak dependence of the energy on the magnetic order
along the $c$ axis was also reported in Ref.~\onlinecite{OKZG+09}. This
nearly two-dimensional character of the magnetic interactions is not
affected by F doping.

\begin{figure}[tbp]
\begin{center}
\includegraphics[width=0.48\textwidth]{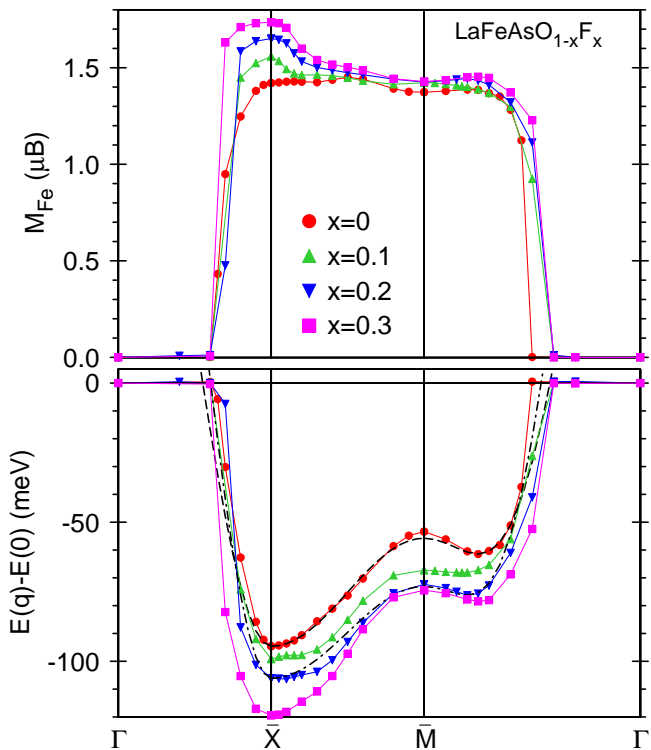}
\end{center}
\caption{(Color online) Same as Fig.~\protect\ref{fig:emq_LaBa}, but for
LaFeAsO$_{1-x}$F$_{x}$ without downshifting the Fe $d_{xy}$ energies.}
\label{fig:emq_La_unsh}
\end{figure}

The $\mathbf{q}$ dependencies of the spin-spiral moment and energy
calculated for LaFeAsO$_{1-x}$F$_{x}$ without downshifting the $d_{xy}$ 
energies are shown in Fig.~\ref{fig:emq_La_unsh}. Also in this case is the
energy minimum of undoped LaFeAsO at $\bar{\mathrm{X}}$, but $\sim 20$ meV
deeper. The local minimum along \={M}$\bar{\Gamma}$ is at the same
$\mathbf{q}$ position, but merely 10 meV deeper, so that the fitted values,
$j_{1}=93$ 
and $j_{2}=66$ meV, are a bit larger but have the same ratio. For
electron-doped LaFeAsO$_{1-x}$F$_{x}$, the $\mathbf{q}$ dependence of the
energy is somewhat stronger and exhibits a shoulder at $\mathbf{q}\approx
(1,0.3)$ which does \emph{not} develop into a well-defined minimum.
Nevertheless, destabilization of the commensurate collinear stripe order 
\emph{does} occur, although the details are seen to depend sensitively on
the underlying band structure.

\begin{table}[tbp]
  \caption{The doping dependence of the Fe magnetic moments
    $M(\bar{\mathrm{X}})$ calculated for stripe AF and the exchange coupling
    constants $j_1$ and $j_2$ in LaFeAsO$_{1-x}$F$_{x}$ and
    Ba$_{1-2y}$K$_{2y}$Fe$_{2}$As$_{2}$. For LaFeAsO$_{1-x}$F$_{x}$, the values
    calculated both with and without the shift of the $d_{xy}$ states are
    presented. The latter are given in parentheses.}  
\label{tab:j}%
\begin{ruledtabular}
\begin{tabular}{cccccc}
 Compound &  $x$/$y$ & $M(\bar{\mathrm{X}})$ ($\mu_{\text{B}}$) &  $j_2/j_1$ 
& \multicolumn{1}{c}{$j_1$ (meV)} & $j_2$ (meV) \\
\hline
LaFeAsO$_{1-x}$F$_{x}$ & 0.3 &  1.68  &  0.71  &  61  &  43  \\
       &     & (1.74) & (0.71) &( 77) & (54) \\
       & 0.2 &  1.58  &  0.71  &  47  &  33  \\
       &     & (1.65) & (0.62) &(103) & (63) \\
       & 0.1 &  1.47  &  0.66  &  71  &  47  \\ 
       &     & (1.56) & (0.59) &(160) & (94) \\
       &  0  &  1.34  &  0.71  &  81  &  57  \\ 
       &     & (1.42) & (0.71) &( 93) & (66) \\
\hline
Ba$_{1-2y}$K$_{2y}$Fe$_{2}$As$_{2}$  &  0  &  1.34  &  0.77  &  95  &  73  \\
       & 0.1 &  1.19  &  0.85  &  77  &  65  \\
       & 0.2 &  1.11  &  1.10  &  39  &  43  \\
       & 0.3 &  0.99  &  1.10  &  19  &  21  \\
\end{tabular}
\end{ruledtabular}
\end{table}

\subsection{\label{sec:sba}Pure and hole-doped Ba$_{1-2y}$K$_{2y}$Fe$_{2}$As$_{2}$}

The spin-spiral moments and energies as functions of $\mathbf{q}=\left(
q_{x},q_{y},0\right) $ for Ba$_{1-2y}$K$_{2y}$Fe$_{2}$As$_{2}$ with 
$y$=0, 0.1, 0.2, and 0.3 holes per FeAs are shown in the right-hand panel of
Fig.~\ref{fig:emq_LaBa}. For undoped BaFe$_{2}$As$_{2}$, these curves (red
dots) are qualitatively similar to those calculated for LaFeAsO, and the
trend for increasing hole-doping of Ba$_{1-2y}$K$_{2y}$Fe$_{2}$As$_{2}$
continues the trend for decreasing electron-doping of LaFeAsO$_{1-x}$F$_{x}$,
i.e., $y\sim -x$. The calculated energy gain $E(\bar{\Gamma})-E(\bar{%
\mathrm{X}})$ due to the formation of stripe AFM order in BaFe$_{2}$As$_{2}$
is somewhat smaller than in LaFeAsO, but the energy difference $E_{\min
}\left( \mathrm{\bar{M}}\bar{\Gamma}\right) -E\left( \mathrm{\bar{X}}\right) 
$ is a bit larger and this causes the fitted values $j_{1}=95$ and $j_{2}=73$
meV to be a bit larger. The ratio $j_{2}/j_{1}$=0.77 is slightly larger than
for LaFeAsO. Hole doping strongly reduces the stripe-formation energy, $E(%
\bar{\Gamma})-E(\bar{\mathrm{X}})$. Nevertheless, the energy minimum remains
at $\bar{\mathrm{X}}$ until the hole doping exceeds 25\%, at which point
the minimum splits in two with the lowest lying along \={X}$\bar{\Gamma}$. Our
calculations thus show that stripe order is more resistant to hole doping
in Ba$_{1-2y}$K$_{2y}$Fe$_{2}$As$_{2}$ than to electron doping in LaFeAsO$%
_{1-x}$F$_{x}$. This conclusion is supported by experimental observations of
the traces of the spin-density-wave phase for K-doping as high as 0.4, i.e.,
well into the superconducting region.\cite{CRQB+08,x:GABB+08}

The local minimum along $\bar{\mathrm{M}}\bar{\Gamma}$ moves toward $\bar{%
\Gamma}$ as hole doping increases. As a consequence, the estimated 
$j_{2}/j_{1}$ increases and reaches the value of 1.1 when $y=0.2$. Since the
energy difference $E_{\min }\left( \mathrm{\bar{M}}\bar{\Gamma}\right)
-E\left( \mathrm{\bar{X}}\right) $ decreases with hole doping, the values of
the effective coupling constants decrease to $j_{1}=39$ and 43 meV when 
$y=0.2$.

The Fe moment calculated for stripe AFM order decreases from 1.34 $\mu _{%
\text{B}}$ in the undoped compound to 0.99 $\mu _{\text{B}}$ for $y=0.3$.
The maximum of $M(\mathbf{q})$ is however not at $\bar{\mathrm{X}}$, but at 
$\mathbf{q}\approx (1,0.4)$ along $\bar{\mathrm{X}}\bar{\mathrm{M}}$. This
maximum becomes more pronounced with hole doping. In contrast to the
situation in LaFeAsO$_{1-x}$F$_{x}$, where for $\mathbf{q}$ in a large
region around $\bar{\Gamma}$ the non-magnetic solution is stable, in Ba$%
_{1-2y}$K$_{2y}$Fe$_{2}$As$_{2}$ magnetic solutions exist closer and closer
to $\bar{\Gamma}$ and with increasing moment as the hole doping increases.
For $y=0.3$ the moment is nearly 0.5 $\mu _{B}$, except very close to $\bar{%
\Gamma}$. A FM solution is thus being approached.

Calculations for spirals with non-zero $q_{z}$ reveal much stronger
dependence of the energy on the relative orientation of the Fe moments in
adjacent FeAs layers than in LaFeAsO. This is due to the stronger $k_{z}$
dispersion of the As-$p_{z}$ hybridized bands which was discussed in
Sec.~\ref{sec:lsda}. In BaFe$_{2}$As$_{2}$, inter-layer nearest As neighbors 
are on top of each other and we find that the lowest-energy solution is for
nearest-neighbor layers having parallel AFM-ordered stripes. The energy for
FM ordering between parallel stripes is 4 meV/Fe higher, and the energies
for orthogonal stripes are intermediate. This is in accord with the
experimental observations \cite{HQBG+08} and results obtained from
calculations for collinear spin arrangements. \cite{Singh08_Ba}

\begin{figure}[tbp]
\begin{center}
\includegraphics[width=0.48\textwidth]{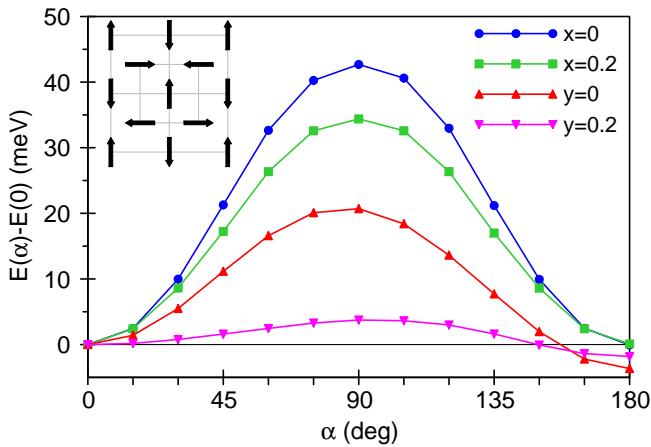}
\end{center}
\caption{(Color online) Dependences of the energy of LaFeAsO$_{1-x}$F$_{x}$
and Ba$_{1-2y}$K$_{2y}$Fe$_{2}$As$_{2}$ on the angle $\protect\alpha$
between the moments of Fe nearest neighbors. The inset (turned 45$^\circ$
with respect to the one in Fig. \protect\ref{fig:emq_LaBa}) shows the spin
arrangement in the Fe plane for $\protect\alpha =90{}^\circ$}
\label{fig:Ea}
\end{figure}

\subsection{Applicability of the simple $j_{1}$-$j_{2}$ Heisenberg model}

The values of the stripe-ordered moment, $M\left( \mathrm{\bar{X}}\right)$,
the $j_{2}/j_{1}$ ratio determined by the position of the local minimum
along $\bar{\mathrm{M}}\bar{\Gamma}$, as well as the values of $j_{1}$ and 
$j_{2}$ are collected in Table \ref{tab:j} for all electron and hole dopings
considered. Whereas $M\left( \mathrm{\bar{X}}\right) $ and $j_{2}/j_{1}$
exhibit clear trends with doping and position of the $d_{xy}$ band, $j_{1}$
and $j_{2}$ scatter more. This enforces a conclusion that an
incommensurable SDW emerging upon electron doping is a typical band-structure
effect: The magnetic energies of the doped compounds can hardly be described
by the simple $j_{1}$--$j_{2}$ Heisenberg model.

In order to test the applicability of $j_{1}$--$j_{2}$ Heisenberg model for
merely the parent compounds, we then performed calculations for spin
structures with $\mathbf{q}=(1,0)$, but with the phases $\varphi_{i}$ for the
two Fe sites in the tetragonal unit cell chosen in such a way that the angle,
$\alpha=\varphi_{1}-\varphi_{2}$,
between their magnetic moments varied from 0 to 180$^{\circ }$. As seen from
the inset in Fig.~\ref{fig:Ea}, which is turned 45${}^\circ$
with respect to the one in Fig. \ref{fig:emq_LaBa}, the Fe sites form two
interpenetrating square sublattices. At $\mathbf{q}=(1,0)$, Fe moments in
each sublattice are ordered antiferromagnetically as shown in the inset. The
angles between the moment of an Fe and those on the two pairs of nearest
neighbors, belonging to the other sublattice, are respectively $\alpha $ and 
$\pi -\alpha $ and, hence, the Heisenberg energy is independent of $\alpha$. 
The calculated dependencies of the energy on $\alpha $ are shown in Fig.~%
\ref{fig:Ea}. The spin arrangement corresponding to $\alpha $=180$^{\circ }$
is exactly the same as the one generated by the spiral with $\mathbf{q}%
=(0,1) $ and $\alpha $=0. In other words, the spirals with $\alpha $=0 and 
$\alpha $=180$^{\circ }$ result in stripe order with the AFM chains parallel
to $x$ and $y$ axis, respectively. The corresponding solutions for LaFeAsO 
are degenerate. In BaFe$_{2}$As$_{2}$ with the body-centered unit cell the
variation in $\alpha $ is accompanied by a change of angle between the Fe
moments in the adjacent FeAs layers. Due to the AFM alignment of the Fe
moments along the $c$ axis, the solution with $\alpha $=180$^{\circ }$ has a
lower energy than that with $\alpha $=0 having FM order along $c$.

Our LSDA calculations show that for both compounds the energy depends
strongly on the relative orientation of the Fe moments, with $E(\alpha 
\mathrm{=}90)-E(\alpha \mathrm{=}0)$ being comparable to the energy
difference between the two collinear AFM solutions with $\mathbf{q}=(1,0)$
and $\mathbf{q}=(1,1)$ discussed in connection with Fig. \ref{fig:emq_LaBa}.
The calculated results behave like $E(\alpha )=C\sin ^{2}\alpha $, which
does not appear in the $j_{1}$--$j_{2}$ Heisenberg model but can be
recovered by adding a biquadratic term proportional to $(\mathbf{S}_{i}\cdot 
\mathbf{S}_{j})^{2}$, with $\mathbf{S}_{i}$ and $\mathbf{S}_{j}$ being spins
on Fe nearest neighbors.

The strong dependence of the energy on the relative orientation of the two
AFM Fe sublattices points to the non-Heisenberg character of the
interactions between Fe moments even in undoped LaFeAsO and BaFe$_{2}$As$%
_{2}$. The $E(\alpha )$ curves calculated for doped compounds (Fig.~\ref%
{fig:Ea}) show a similar, although weaker, $\alpha$ dependence,
especially for Ba$_{0.6}$K$_{0.4}$Fe$_{2}$As$_{2}$.

Thus, among the variety of spin configurations, which were degenerate in the 
$j_{1}$--$j_{2}$ Heisenberg model, a collinear spin arrangement with either 
$\mathbf{q}=(1,0)$ or $\mathbf{q}=(0,1)$ is favored already at the level of
the LSDA electronic structure. Such a magnetically ordered solutions lower
the symmetry of the lattice from tetragonal to orthorhombic and lifts the
degeneracy of Fe $d_{yz}$ and $d_{zx}$ states. This symmetry lowering is
apparently responsible for the anisotropy of the exchange interactions
calculated for stripe AFM order in Ref.~\onlinecite{YLHN+08}.

\section{\label{sec:suscept}Noninteracting susceptibility}

With an Fe moment of $\sim $1.5 $\mu _{\text{B}}$, the exchange splitting
between the majority- and minority-spin Fe $d$ states in the LSDA
calculations is about 1.3 eV, which is as large as the dispersion of the
paramagnetic Fe $d_{xy}$ and $d_{yz}$ and $d_{xz}$ bands over a
significant part of the BZ, e.g., the $d_{yz}$ band in a tube around \={M}\={Y}.
Such a strong magnetic perturbation dramatically changes the band
structure and the topology of the Fermi surface. The limit of a weak
magnetic perturbation can be analyzed by studying the $\mathbf{q}$ and
doping dependencies of the static, noninteracting linear-response
susceptibility. Its imaginary part, $\mathrm{Im}\chi _{0}(\mathbf{q})=%
\mathrm{Im}\chi _{0}(\mathbf{q},\omega \rightarrow 0)/\omega $, is
determined by the shape of and velocities on the Fermi surface (FS) and is a
quantitative measure of FS nesting. The real part, $\mathrm{Re}\chi _{0}(%
\mathbf{q})=\mathrm{Re}\chi _{0}(\mathbf{q},\omega =0)$, describes the
response of the system to an infinitesimally small perturbation. In contrast
to $\mathrm{Im}\chi _{0}(\mathbf{q})$, the electronic states in a wide
energy range around $E_{F}$ may contribute to $\mathrm{Re}\chi _{0}(\mathbf{%
q})$.

In the present work the noninteracting susceptibility 
\begin{eqnarray}
\chi _{0}(\mathbf{q},\omega ) =-\frac{1}{V}\sum_{\mathbf{k},n,n^{\prime }}%
\frac{f_{n^{\prime }}(\mathbf{k}+\mathbf{q})-f_{n}(\mathbf{k})}{\varepsilon
_{n^{\prime }}(\mathbf{k}+\mathbf{q})-\varepsilon _{n}(\mathbf{k})+\omega
+i\delta }  \notag \\
\times \langle \mathbf{k},n|e^{-i\mathbf{q\cdot }\mathbf{r}}|\mathbf{k}+%
\mathbf{q},n^{\prime }\rangle \langle \mathbf{k}+\mathbf{q},n^{\prime }|e^{i%
\mathbf{q\cdot }\mathbf{r}}|\mathbf{k},n\rangle  \label{eq:xi}
\end{eqnarray}
was calculated using the linear response expressions given in Ref.\
\onlinecite{CW75}. Here, $\varepsilon _{n}(\mathbf{k})$ is the energy of the
$n$th 
band and $f_{n}(\mathbf{k})$ is the Fermi function. First, the imaginary
part of $\chi _{0}(\mathbf{q},\omega )$ was calculated in the $\delta
\rightarrow 0$ limit. Then, the real part was obtained by using
Kramers-Kronig relations. The matrix elements $\langle \mathbf{k}+\mathbf{q}%
,n^{\prime }|\exp (i\mathbf{q}\mathbf{r})|\mathbf{k},n\rangle $ of the
perturbation were approximated by expanding the exponent inside each Fe
sphere in Bessel functions and keeping only the spherically symmetric term
proportional to $j_{0}(\mathbf{q}\mathbf{r})$. The expressions for the
matrix elements were further simplified by using $j_{0}(\mathbf{q}\mathbf{r}%
)\approx 1$. Within this approximation the contribution of each Fe sphere to
the matrix element $\langle \mathbf{k}+\mathbf{q},n^{\prime }|\exp (i\mathbf{%
q}\mathbf{r})|\mathbf{k},n\rangle $ is proportional to the overlap integral
of the LMTO wave functions $\Psi _{\mathbf{k},n}$ and $\Psi _{\mathbf{k}+%
\mathbf{q},n^{\prime }}$ inside the sphere. In other words, two states
contribute to the susceptibility only if they have similar Fe partial-wave
character. Although these approximations are valid only at sufficiently
small $|\mathbf{q}|$, they do not affect the analysis of susceptibility
peaks which may appear due to the FS nesting. More details on calculation of 
$\chi _{0}(\mathbf{q},\omega )$ using the LMTO method can be found in Ref.~%
\onlinecite{YYFT07}.

The contribution of a particular subset of electronic states to $\chi _{0}(%
\mathbf{q},\omega )$ can be discerned by calculating the matrix elements of
the perturbation with all coefficients of the LMTO wave functions, except
those which correspond to the chosen subset, set to zero. Due to the
presence of interference terms, such orbitally resolved contributions to the
susceptibility are not additive. However, they allow us to discriminate
those states which give the dominant contribution to $\chi _{0}(\mathbf{q}%
,\omega )$.

The susceptibilities of iron pnictides presented below were calculated
starting from self-consistent non-spin-polarized electron densities and
neglecting spin-orbit coupling. Thus, because of the degeneracy of Bloch
states with different spin projections, $\chi _{0}(\mathbf{q},\omega )$ 
calculated using Eq.~(\ref{eq:xi}) describes the response of the system to
a $\mathbf{q}$-dependent spin as well as charge perturbation. A 32$\times $%
32$\times $32 $\mathbf{k}$ mesh in the small tetragonal BZ was used in the
calculations.

\subsection{\label{sec:xla}LaFeAsO$_{1-x}$F$_x$}

Nesting between the quasi-two-dimensional FS sheets in the iron pnictides
were noted already in the earliest electronic-structure calculations.\cite%
{MSJD08,DZXL+08,OKZG+09} The FS cross sections calculated for undoped
LaFeAsO with and without the downwards shift of the Fe $d_{xy}$ energy
were shown in Fig.~\ref{fig:La_fs} and their nesting pointed out in the
caption to this figure, as well as in Sec.~\ref{sec:lsda}.

\begin{figure*}[tbp]
\begin{center}
\includegraphics[width=0.96\textwidth]{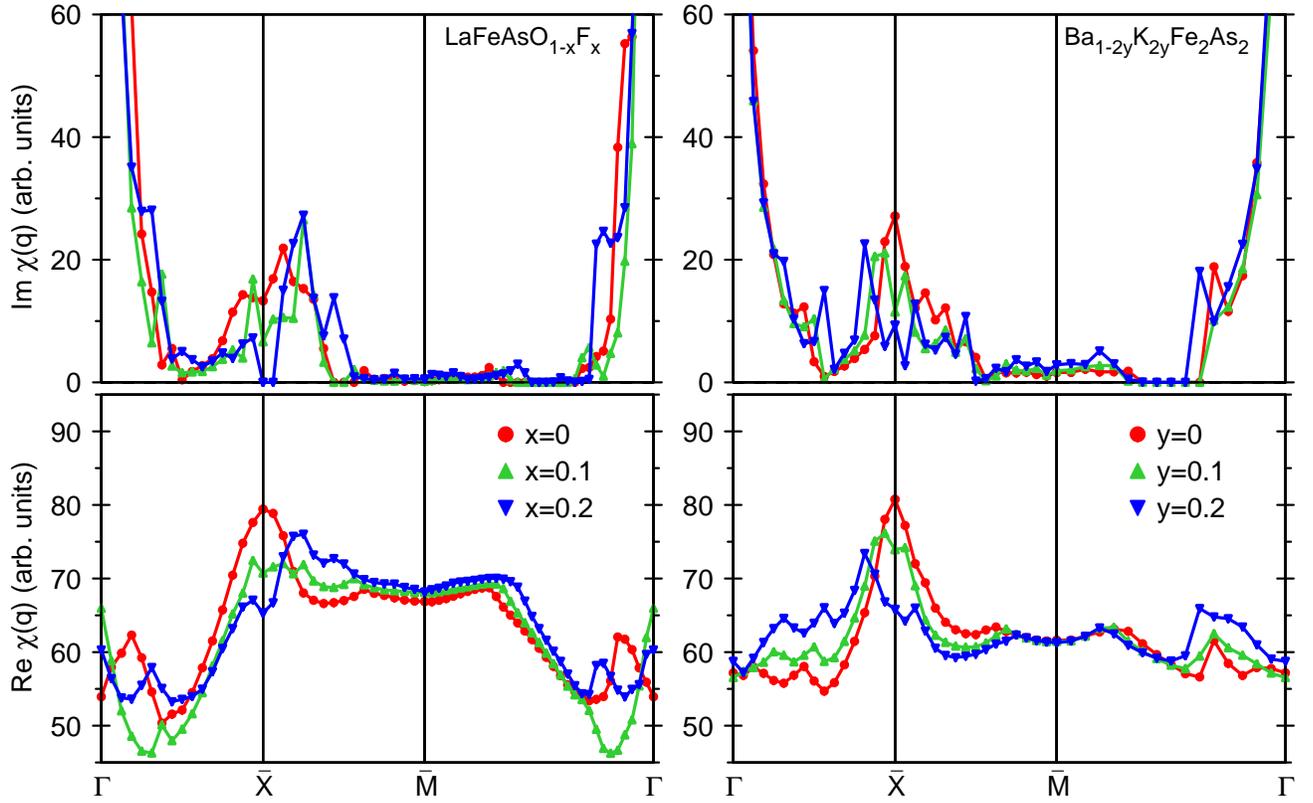}
\end{center}
\caption{(Color online) Imaginary and real parts of the bare static
susceptibility for LaFeAsO$_{1-x}$F$_{x}$ and
Ba$_{1-2y}$K$_{2y}$Fe$_{2}$As$_{2}$.} 
\label{fig:xiq_LaBa}
\end{figure*}

The imaginary and real parts of the bare static susceptibility calculated for
LaFeAsO$_{1-x}$F$_{x}$ 
with downshifted $d_{xy}$ energy are shown in the left-hand panel of Fig.~%
\ref{fig:xiq_LaBa}. We first discuss the results for the undoped $\left(
x=0\right) $ compound. In agreement with previous results \cite%
{MSJD08,DZXL+08} the maximum of $\mathrm{Re}\chi _{0}(\mathbf{q})$ is found
at $\mathbf{q}=\mathrm{\bar{X}}\left( 1,0\right) $. A peak of the imaginary
part is, however, shifted away from $\bar{\mathrm{X}}$ toward
$\bar{\mathrm{M}}$.  
Analysis of the partial-wave resolved contributions to the susceptibility
shows that the main contribution to $\mathrm{Im}\chi _{0}(\mathrm{\bar{X}})$
comes from nesting of \={M}-centered $d_{yz,xz}$-like hole sheet (blue in
the left-hand side of Fig.~\ref{fig:La_fs}) with the \={Y}-centered
electron sheet (green). This also gives the dominant contribution to the
maximum of $\mathrm{Re}\chi _{0}(\mathrm{\bar{X}})$. But also the $d_{xy}$ 
states (red) contribute significantly to $\mathrm{Re}\chi _{0}(\bar{\mathrm{X%
}})$ although their contribution to $\mathrm{Im}\chi _{0}(\mathrm{\bar{X}})$
nearly vanishes due to the ineffective nesting of the innermost
$\bar{\Gamma}$-centered hole sheet with the \={X}-centered electron sheet. As
$\mathbf{q}$ 
moves along $\bar{\mathrm{X}}\bar{\mathrm{M}}$, the hole and electron sheets
start to touch when $\mathbf{q}\approx \left( 1,0.13\right) $ and a peak of
the $d_{xy}$ contribution to $\mathrm{Im}\chi _{0}(\mathbf{q})$ appears at
this nesting vector. This is responsible for the maximum of $\mathrm{Im}\chi
_{0}(\mathbf{q})$ along $\bar{\mathrm{X}}\bar{\mathrm{M}}$. The $d_{xy}$ 
contribution to $\mathrm{Re}\chi _{0}(\mathbf{q})$ reaches its maximum at
the same $\mathbf{q}$. Since the weight of the Fe $d_{3z^{2}-1}$ and 
$d_{x^{2}-y^{2}}$ states in the bands crossing $E_{F}$ is very small, they
do not contribute to $\mathrm{Im}\chi _{0}(\mathbf{q})$, whereas their
contribution to $\mathrm{Re}\chi _{0}(\mathbf{q})$ is almost constant in
the whole $\mathbf{q}$ range.

With electron doping, the hole sheets, centered at $\bar{\Gamma}$ and \={M},
shrink and the electron sheets, centered at \={X} and \={Y}, grow. The hole
and electron sheets no longer nest for $\mathbf{q}=$\={X} so that the
susceptibility at $\bar{\mathrm{X}}$ decreases rapidly with doping. Instead,
peaks develop in both the imaginary and real parts of $\chi _{0}(\mathbf{q})$
 for $\mathbf{q}$ along $\bar{\mathrm{X}}\bar{\mathrm{M}}$ for which the
electron and hole sheets touch. This shift of the susceptibility peaks with
increasing $x$ correlates with the shift of the minimum of the $E(\mathbf{q})$
 curves calculated for spin spirals (Fig.~\ref{fig:emq_LaBa}). The shift
of the $\mathrm{Re}\chi _{0}(\mathbf{q})$ peak along the $\bar{\mathrm{X}}%
\bar{\Gamma}$ line, accompanied by strong suppression at \={X}, was noted in
Ref.~\onlinecite{DZXL+08}.

\begin{figure}[tbp]
\begin{center}
\includegraphics[width=0.48\textwidth]{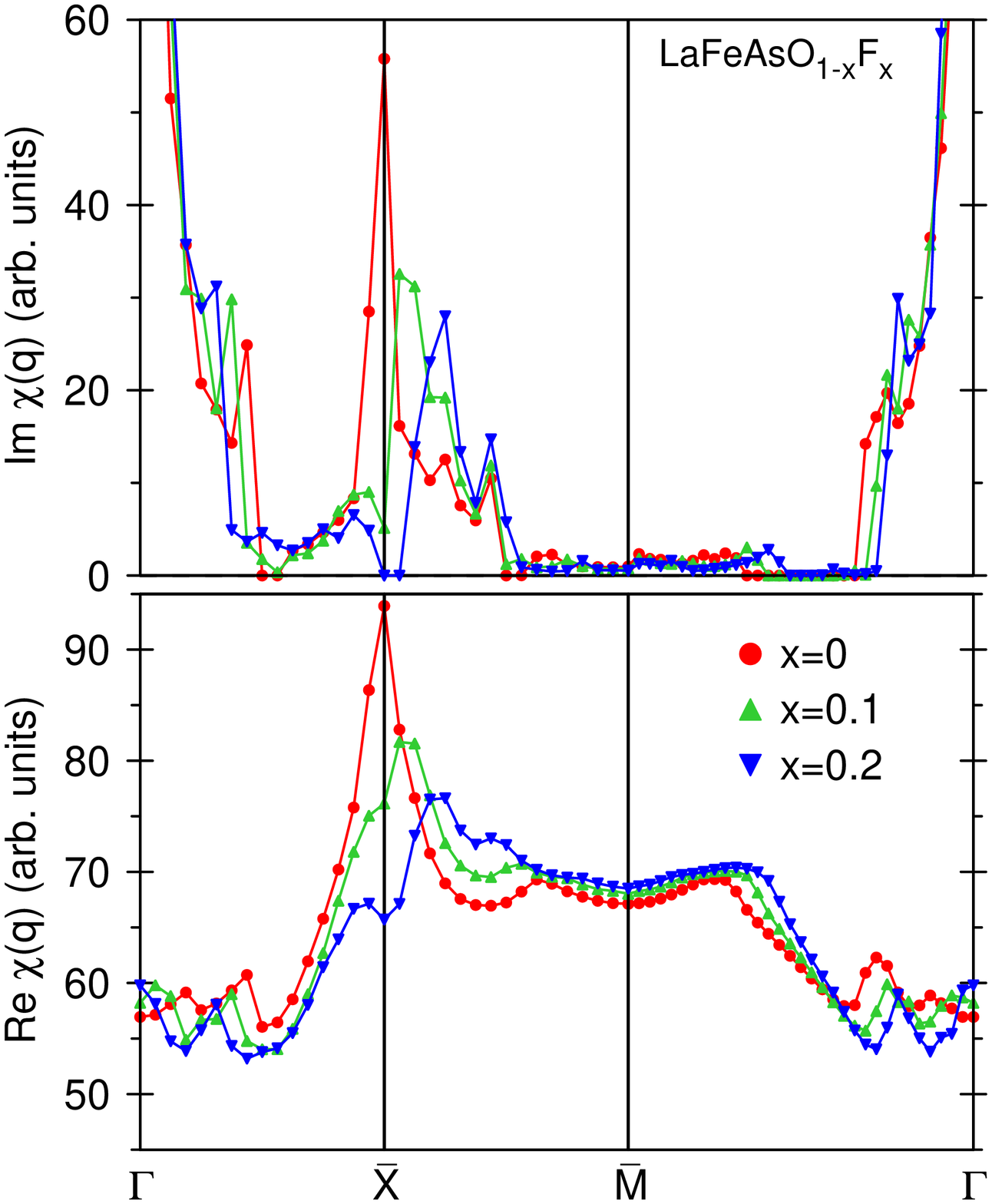}
\end{center}
\caption{(Color online) Imaginary and real parts of the bare susceptibility
calculated for LaFeAsO$_{1-x}$F$_{x}$ without shifting of the $d_{xy}$ 
energy.}
\label{fig:xiq_La_unsh}
\end{figure}

Comparison of $\chi _{0}(\mathbf{q})$ calculated with (Fig.~\ref%
{fig:xiq_LaBa}) and without (Fig.~\ref{fig:xiq_La_unsh}) shifting the Fe 
$d_{xy}$ energy shows that the susceptibility of undoped LaFeAsO is
sensitive to the changes of the FS shown in Fig.~\ref{fig:La_fs} caused by
the shift. Without the shift, the $\bar{\Gamma}$-centered $d_{xy}$-like hole
sheet is larger and nests almost perfectly the \={X}-centered electron sheet
when $\mathbf{q}=\mathrm{\bar{X}}$. This leads to $\mathrm{Im}\chi _{0}(%
\mathrm{\bar{X}})$ being the top of a sharp peak with the dominant
contribution coming from the $d_{xy}$ states. The $d_{xy}$ contribution to 
$\mathrm{Re}\chi _{0}(\mathbf{q})$ also has a sharp maximum at $\bar{%
\mathrm{X}}$. Since even the largest of the two \={M}-centered $d_{yz,zx}$ 
hole sheets is smaller than the $\bar{\mathrm{Y}}$-centered electron sheet,
nesting of these sheets does not contribute to $\mathrm{Im}\chi _{0}(\bar{%
\mathrm{X}})$. Their contribution increases as $\mathbf{q}$ shifts away
from $\bar{\mathrm{X}}$, but remains weaker than the $d_{xy}$ one.
Nevertheless, the $d_{yz,zx}$ contribution to $\mathrm{Re}\chi _{0}(\mathbf{%
q})$ in the vicinity of the $\bar{\mathrm{X}}$ point is comparable to the 
$d_{xy}$ one. The $\mathbf{q}$ dependence of the imaginary part of
susceptibility is strongly affected by the change of FS nesting caused by
the shift of the $d_{xy}$ bands, but the behavior of the real part is much
more robust. Its maximum does not move away from the $\bar{\mathrm{X}}$ 
point but looses its sharpness as the FS nesting becomes less perfect.

The doping dependence of the susceptibility is only weakly affected by the
shift of the $d_{xy}$ states. This may be one reason why different layered
iron arsenides exhibit similar properties, in spite of the variation in
their band structures.

\subsection{\label{sec:xba}Ba$_{1-2y}$K$_{2y}$Fe$_2$As$_2$}

As explained in Sec.~\ref{sec:lsda}, the FS of BaFe$_{2}$As$_{2}$ is
similar to that of LaFeAsO but all cylinders, except the $\bar{\Gamma}$%
-centered $d_{xy}$-like hole cylinder, are far more warped in $k_{z}$
direction. Besides, the bct symmetry mixes $k_{z}$ dispersion into the $%
\left( k_{x},k_{y}\right)$ dispersions. Although the nesting for $q=\mathrm{%
\bar{X}}$ is good when $k_{z}=\pi /2c$, it deteriorates at other $k_{z}$.
Nevertheless, a peak of $\mathrm{Im}\chi _{0}(\mathbf{q})$, mostly due to
the $d_{xy}$ states, is still present at $\bar{\mathrm{X}}$ (right-hand
side of Fig.~\ref{fig:emq_LaBa}), and the behavior of $\mathrm{Re}\chi _{0}(%
\mathbf{q})$ for $y=0$ is qualitatively similar to that for LaFeAsO. This
comparison leads to a conclusion that independently of the fine details of
FS nesting, the formation of commensurate stripe order in the undoped
compounds is preferable also in the limit of a weak magnetism.

Hole doping makes the hole sheets grow and the electron sheets shrink. The
nesting with $\mathbf{q}=$\={X} is, nonetheless, destroyed just as
effectively as by electron doping. As a consequence, the real and imaginary
parts of the susceptibility at $\bar{\mathrm{X}}$ are strongly suppressed.
In contrast to LaFeAsO$_{1-x}$F$_{x}$, the peak in $\mathrm{Re}\chi _{0}(%
\mathbf{q})$ shifts toward $\bar{\Gamma}$, like the energy minimum
calculated for spin spirals in Ba$_{1-2y}$K$_{2y}$Fe$_{2}$As$_{2}$ (Fig.~%
\ref{fig:emq_LaBa}) for the highest doping. The appearance of small-$q$,
small-moment spin spirals correlates with the flat nonvanishing behavior
of $\mathrm{Re}\chi _{0}(\mathbf{q})$ near $\bar{\Gamma}$.

\section{\label{sec:concl}Conclusions}

In conclusion, our LSDA total-energy calculations for spin spirals in LaFeAsO%
$_{1-x}$F$_{x}$ and Ba$_{1-2y}$K$_{2y}$Fe$_{2}$As$_{2}$ using the
experimental crystal structures confirm that in the undoped compounds the
minimum of the total energy is reached at the wave vector $\mathbf{q}=%
\mathrm{\bar{X}}(1,0)$, which corresponds to stripe AFM order. The stability
of this solution is, however, strongly affected by doping. With
electron doping $\left( x\right) $ exceeding 0.1 in LaFeAsO$_{1-x}$F$_{x}$,
the minimum becomes shallow and shifts toward \={M}$\left( 1,1\right) $ to
an incommensurate wave vector. 
This destabilization of the commensurate collinear stripe order by electron
doping is a band-structure effect but not sensitive to the details of Fermi
surface nesting. Hole doping $\left( y\right) $ in Ba$_{1-2y}$K$_{2y}$Fe$_{2}
$As$_{2}$ roughly continues the trend calculated for decreasing electron
doping in LaFeAsO$_{1-x}$F$_{x}$, that is, $y\sim -x$. The energy gain due
to stripe formation decreases with hole doping, but the minimum stays at
$\bar{\mathrm{X}}$ for $y\lesssim 0.25$. In both compounds, the deviation of
the $\mathbf{q}$ dependence of the energy from that of the classical
Heisenberg model with nearest- and next-nearest-neighbor interactions becomes
more pronounced with doping.

We found that even in the parent compounds the total energy varies strongly
when two interpenetrating AFM sublattices formed by the Fe ions are rotated
with respect to each other. The dependence of the energy on the angle
between the Fe moments in the two sublattices cannot be described by the
simple $j_{1}$--$j_{2}$ Heisenberg model, but may be reproduced by a
biquadratic term proportional to $(\mathbf{S}_{i}\cdot \mathbf{S}_{j})^{2}$,
which favors collinear stripe AFM order.

Although the LSDA for the experimental crystal structures gives a stripe
moments around 1.5 $\mu _{B}/$Fe and concomitant eV-large exchange
splittings of degenerate $\varepsilon _{j,\mathbf{k}}$ and $\varepsilon
_{j^{\prime },\mathbf{k}+\mathbf{q}}$ bands, linear-response calculations of
the real and imaginary parts of the static noninteracting susceptibility
based on the paramagnetic LDA band structure give similar results as the
charge- and spin-self-consistent spin-spiral calculations.

Authors are grateful to G.~Jackeli and L.~Boeri for helpful discussions.
V.N. Antonov gratefully acknowledges the hospitality at Max-Planck-Institut
f\"{u}r Festk\"{o}rperforschung in Stuttgart during his stay there. This work
was partially supported by Science and Technology Center in Ukraine (STCU)
inder Project No.~4930.

\bibliography{./jprb,./XFeAs,./g,./yar}

\newcommand{\noopsort}[1]{} \newcommand{\printfirst}[2]{#1}
  \newcommand{\singleletter}[1]{#1} \newcommand{\switchargs}[2]{#2#1}
\begin{thebibliography}{35}
\expandafter\ifx\csname natexlab\endcsname\relax\def\natexlab#1{#1}\fi
\expandafter\ifx\csname bibnamefont\endcsname\relax
  \def\bibnamefont#1{#1}\fi
\expandafter\ifx\csname bibfnamefont\endcsname\relax
  \def\bibfnamefont#1{#1}\fi
\expandafter\ifx\csname citenamefont\endcsname\relax
  \def\citenamefont#1{#1}\fi
\expandafter\ifx\csname url\endcsname\relax
  \def\url#1{\texttt{#1}}\fi
\expandafter\ifx\csname urlprefix\endcsname\relax\def\urlprefix{URL }\fi
\providecommand{\bibinfo}[2]{#2}
\providecommand{\eprint}[2][]{\url{#2}}

\bibitem[{\citenamefont{Kamihara et~al.}(2008)\citenamefont{Kamihara, Watanabe,
  Hirano, and Hosono}}]{KWHH08}
\bibinfo{author}{\bibfnamefont{Y.}~\bibnamefont{Kamihara}},
  \bibinfo{author}{\bibfnamefont{T.}~\bibnamefont{Watanabe}},
  \bibinfo{author}{\bibfnamefont{M.}~\bibnamefont{Hirano}}, \bibnamefont{and}
  \bibinfo{author}{\bibfnamefont{H.}~\bibnamefont{Hosono}},
  \bibinfo{journal}{J. Am. Chem. Soc.} \textbf{\bibinfo{volume}{130}},
  \bibinfo{pages}{3296} (\bibinfo{year}{2008}).

\bibitem[{\citenamefont{Ren et~al.}(2008{\natexlab{a}})\citenamefont{Ren, Yang,
  Lu, Yi, Che, Dong, Sun, and Zhao}}]{RYLY+08_Pr}
\bibinfo{author}{\bibfnamefont{Z.~A.} \bibnamefont{Ren}},
  \bibinfo{author}{\bibfnamefont{J.}~\bibnamefont{Yang}},
  \bibinfo{author}{\bibfnamefont{W.}~\bibnamefont{Lu}},
  \bibinfo{author}{\bibfnamefont{W.}~\bibnamefont{Yi}},
  \bibinfo{author}{\bibfnamefont{G.~C.} \bibnamefont{Che}},
  \bibinfo{author}{\bibfnamefont{X.~L.} \bibnamefont{Dong}},
  \bibinfo{author}{\bibfnamefont{L.~L.} \bibnamefont{Sun}}, \bibnamefont{and}
  \bibinfo{author}{\bibfnamefont{Z.~X.} \bibnamefont{Zhao}},
  \bibinfo{journal}{Materials Research Innovations}
  \textbf{\bibinfo{volume}{12}}, \bibinfo{pages}{105}
  (\bibinfo{year}{2008}{\natexlab{a}}).

\bibitem[{\citenamefont{Ren et~al.}(2008{\natexlab{b}})\citenamefont{Ren, Yang,
  Lu, Yi, Shen, Li, Che, Dong, Sun, Zhou, and Zhao}}]{RYLY+08_Nd}
\bibinfo{author}{\bibfnamefont{Z.-A.} \bibnamefont{Ren}},
  \bibinfo{author}{\bibfnamefont{J.}~\bibnamefont{Yang}},
  \bibinfo{author}{\bibfnamefont{W.}~\bibnamefont{Lu}},
  \bibinfo{author}{\bibfnamefont{W.}~\bibnamefont{Yi}},
  \bibinfo{author}{\bibfnamefont{X.-L.} \bibnamefont{Shen}},
  \bibinfo{author}{\bibfnamefont{Z.-C.} \bibnamefont{Li}},
  \bibinfo{author}{\bibfnamefont{G.-C.} \bibnamefont{Che}},
  \bibinfo{author}{\bibfnamefont{X.-L.} \bibnamefont{Dong}},
  \bibinfo{author}{\bibfnamefont{L.-L.} \bibnamefont{Sun}},
  \bibinfo{author}{\bibfnamefont{F.}~\bibnamefont{Zhou}}, \bibnamefont{and}
  \bibinfo{author}{\bibfnamefont{Z.-X.} \bibnamefont{Zhao}},
  \bibinfo{journal}{EPL} \textbf{\bibinfo{volume}{82}}, \bibinfo{pages}{57002}
  (\bibinfo{year}{2008}{\natexlab{b}}).

\bibitem[{\citenamefont{Rotter et~al.}(2008{\natexlab{a}})\citenamefont{Rotter,
  Tegel, and Johrendt}}]{RTJ08}
\bibinfo{author}{\bibfnamefont{M.}~\bibnamefont{Rotter}},
  \bibinfo{author}{\bibfnamefont{M.}~\bibnamefont{Tegel}}, \bibnamefont{and}
  \bibinfo{author}{\bibfnamefont{D.}~\bibnamefont{Johrendt}},
  \bibinfo{journal}{Phys. Rev. Lett.} \textbf{\bibinfo{volume}{101}},
  \bibinfo{pages}{107006} (\bibinfo{year}{2008}{\natexlab{a}}).

\bibitem[{\citenamefont{Ni et~al.}(2008)\citenamefont{Ni, Bud'ko, Kreyssig,
  Nandi, Rustan, Goldman, Gupta, Corbett, Kracher, and Canfield}}]{NBKN+08}
\bibinfo{author}{\bibfnamefont{N.}~\bibnamefont{Ni}},
  \bibinfo{author}{\bibfnamefont{S.~L.} \bibnamefont{Bud'ko}},
  \bibinfo{author}{\bibfnamefont{A.}~\bibnamefont{Kreyssig}},
  \bibinfo{author}{\bibfnamefont{S.}~\bibnamefont{Nandi}},
  \bibinfo{author}{\bibfnamefont{G.~E.} \bibnamefont{Rustan}},
  \bibinfo{author}{\bibfnamefont{A.~I.} \bibnamefont{Goldman}},
  \bibinfo{author}{\bibfnamefont{S.}~\bibnamefont{Gupta}},
  \bibinfo{author}{\bibfnamefont{J.~D.} \bibnamefont{Corbett}},
  \bibinfo{author}{\bibfnamefont{A.}~\bibnamefont{Kracher}}, \bibnamefont{and}
  \bibinfo{author}{\bibfnamefont{P.~C.} \bibnamefont{Canfield}},
  \bibinfo{journal}{Phys. Rev. B} \textbf{\bibinfo{volume}{78}},
  \bibinfo{pages}{014507} (\bibinfo{year}{2008}).

\bibitem[{\citenamefont{Rotter et~al.}(2008{\natexlab{b}})\citenamefont{Rotter,
  Tegel, Johrendt, Schellenberg, Hermes, and Poettgen}}]{RTJS+08}
\bibinfo{author}{\bibfnamefont{M.}~\bibnamefont{Rotter}},
  \bibinfo{author}{\bibfnamefont{M.}~\bibnamefont{Tegel}},
  \bibinfo{author}{\bibfnamefont{D.}~\bibnamefont{Johrendt}},
  \bibinfo{author}{\bibfnamefont{I.}~\bibnamefont{Schellenberg}},
  \bibinfo{author}{\bibfnamefont{W.}~\bibnamefont{Hermes}}, \bibnamefont{and}
  \bibinfo{author}{\bibfnamefont{R.}~\bibnamefont{Poettgen}},
  \bibinfo{journal}{Phys. Rev. B} \textbf{\bibinfo{volume}{78}},
  \bibinfo{pages}{020503(R)} (\bibinfo{year}{2008}{\natexlab{b}}).

\bibitem[{\citenamefont{Nomura et~al.}(2008)\citenamefont{Nomura, Kim,
  Kamihara, Hirano, Sushko, Kato, Takata, Shluger, and Hosono}}]{NKKH+08}
\bibinfo{author}{\bibfnamefont{T.}~\bibnamefont{Nomura}},
  \bibinfo{author}{\bibfnamefont{S.~W.} \bibnamefont{Kim}},
  \bibinfo{author}{\bibfnamefont{Y.}~\bibnamefont{Kamihara}},
  \bibinfo{author}{\bibfnamefont{M.}~\bibnamefont{Hirano}},
  \bibinfo{author}{\bibfnamefont{P.~V.} \bibnamefont{Sushko}},
  \bibinfo{author}{\bibfnamefont{K.}~\bibnamefont{Kato}},
  \bibinfo{author}{\bibfnamefont{M.}~\bibnamefont{Takata}},
  \bibinfo{author}{\bibfnamefont{A.~L.} \bibnamefont{Shluger}},
  \bibnamefont{and} \bibinfo{author}{\bibfnamefont{H.}~\bibnamefont{Hosono}},
  \bibinfo{journal}{Supercon. Sci. Technol.} \textbf{\bibinfo{volume}{21}},
  \bibinfo{pages}{125028} (\bibinfo{year}{2008}).

\bibitem[{\citenamefont{Goko et~al.}(unpublished)\citenamefont{Goko, Aczel,
  Baggio-Saitovitch, Bud'ko, Canfield, Carlo, Chen, Dai, Hamann, Hu, Kageyama,
  Luke, Luo, Nachumi, Ni, Reznik, Sanchez-Candela, Savici, Sikes, Wang, Wiebe,
  Williams, Yamamoto, Yu, and Uemura}}]{x:GABB+08}
\bibinfo{author}{\bibfnamefont{T.}~\bibnamefont{Goko}},
  \bibinfo{author}{\bibfnamefont{A.~A.} \bibnamefont{Aczel}},
  \bibinfo{author}{\bibfnamefont{E.}~\bibnamefont{Baggio-Saitovitch}},
  \bibinfo{author}{\bibfnamefont{S.~L.} \bibnamefont{Bud'ko}},
  \bibinfo{author}{\bibfnamefont{P.~C.} \bibnamefont{Canfield}},
  \bibinfo{author}{\bibfnamefont{J.~P.} \bibnamefont{Carlo}},
  \bibinfo{author}{\bibfnamefont{G.~F.} \bibnamefont{Chen}},
  \bibinfo{author}{\bibfnamefont{P.}~\bibnamefont{Dai}},
  \bibinfo{author}{\bibfnamefont{A.~C.} \bibnamefont{Hamann}},
  \bibinfo{author}{\bibfnamefont{W.~Z.} \bibnamefont{Hu}},
  \bibinfo{author}{\bibfnamefont{H.}~\bibnamefont{Kageyama}},
  \bibinfo{author}{\bibfnamefont{G.~M.} \bibnamefont{Luke}},
  \bibinfo{author}{\bibfnamefont{J.~L.} \bibnamefont{Luo}},
  \bibinfo{author}{\bibfnamefont{B.}~\bibnamefont{Nachumi}},
  \bibinfo{author}{\bibfnamefont{N.}~\bibnamefont{Ni}},
  \bibinfo{author}{\bibfnamefont{D.}~\bibnamefont{Reznik}},
  \bibinfo{author}{\bibfnamefont{D.~R.} \bibnamefont{Sanchez-Candela}},
  \bibinfo{author}{\bibfnamefont{A.~T.} \bibnamefont{Savici}},
  \bibinfo{author}{\bibfnamefont{K.~J.} \bibnamefont{Sikes}},
  \bibinfo{author}{\bibfnamefont{N.~L.} \bibnamefont{Wang}},
  \bibinfo{author}{\bibfnamefont{C.~R.} \bibnamefont{Wiebe}},
  \bibinfo{author}{\bibfnamefont{T.~J.} \bibnamefont{Williams}},
  \bibinfo{author}{\bibfnamefont{T.}~\bibnamefont{Yamamoto}},
  \bibinfo{author}{\bibfnamefont{W.}~\bibnamefont{Yu}}, \bibnamefont{and}
  \bibinfo{author}{\bibfnamefont{Y.~J.} \bibnamefont{Uemura}},
  \bibinfo{journal}{arXiv:0808.1425v1}  (\bibinfo{year}{unpublished}).

\bibitem[{\citenamefont{Huang et~al.}(2008)\citenamefont{Huang, Qiu, Bao,
  Green, Lynn, Gasparovic, Wu, Wu, and Chen}}]{HQBG+08}
\bibinfo{author}{\bibfnamefont{Q.}~\bibnamefont{Huang}},
  \bibinfo{author}{\bibfnamefont{Y.}~\bibnamefont{Qiu}},
  \bibinfo{author}{\bibfnamefont{W.}~\bibnamefont{Bao}},
  \bibinfo{author}{\bibfnamefont{M.~A.} \bibnamefont{Green}},
  \bibinfo{author}{\bibfnamefont{J.~W.} \bibnamefont{Lynn}},
  \bibinfo{author}{\bibfnamefont{Y.~C.} \bibnamefont{Gasparovic}},
  \bibinfo{author}{\bibfnamefont{T.}~\bibnamefont{Wu}},
  \bibinfo{author}{\bibfnamefont{G.}~\bibnamefont{Wu}}, \bibnamefont{and}
  \bibinfo{author}{\bibfnamefont{X.~H.} \bibnamefont{Chen}},
  \bibinfo{journal}{Phys. Rev. Lett.} \textbf{\bibinfo{volume}{101}},
  \bibinfo{pages}{257003} (\bibinfo{year}{2008}).

\bibitem[{\citenamefont{de~la Cruz et~al.}(2008)\citenamefont{de~la Cruz,
  Huang, Lynn, Li, Ratcliff, Zarestky, Mook, Chen, Luo, Wang, and
  Dai}}]{CHLL+08}
\bibinfo{author}{\bibfnamefont{C.}~\bibnamefont{de~la Cruz}},
  \bibinfo{author}{\bibfnamefont{Q.}~\bibnamefont{Huang}},
  \bibinfo{author}{\bibfnamefont{J.~W.} \bibnamefont{Lynn}},
  \bibinfo{author}{\bibfnamefont{J.}~\bibnamefont{Li}},
  \bibinfo{author}{\bibfnamefont{I.}~\bibnamefont{Ratcliff},
  \bibfnamefont{W.}}, \bibinfo{author}{\bibfnamefont{J.~L.}
  \bibnamefont{Zarestky}}, \bibinfo{author}{\bibfnamefont{H.~A.}
  \bibnamefont{Mook}}, \bibinfo{author}{\bibfnamefont{G.~F.}
  \bibnamefont{Chen}}, \bibinfo{author}{\bibfnamefont{J.~L.}
  \bibnamefont{Luo}}, \bibinfo{author}{\bibfnamefont{N.~L.}
  \bibnamefont{Wang}}, \bibnamefont{and}
  \bibinfo{author}{\bibfnamefont{P.}~\bibnamefont{Dai}},
  \bibinfo{journal}{Nature (London)} \textbf{\bibinfo{volume}{453}},
  \bibinfo{pages}{899} (\bibinfo{year}{2008}).

\bibitem[{\citenamefont{Dong et~al.}(2008)\citenamefont{Dong, Zhang, Xu, Li,
  Li, Hu, Wu, Chen, Dai, Luo, Fang, and Wang}}]{DZXL+08}
\bibinfo{author}{\bibfnamefont{J.}~\bibnamefont{Dong}},
  \bibinfo{author}{\bibfnamefont{H.~J.} \bibnamefont{Zhang}},
  \bibinfo{author}{\bibfnamefont{G.}~\bibnamefont{Xu}},
  \bibinfo{author}{\bibfnamefont{Z.}~\bibnamefont{Li}},
  \bibinfo{author}{\bibfnamefont{G.}~\bibnamefont{Li}},
  \bibinfo{author}{\bibfnamefont{W.~Z.} \bibnamefont{Hu}},
  \bibinfo{author}{\bibfnamefont{D.}~\bibnamefont{Wu}},
  \bibinfo{author}{\bibfnamefont{G.~F.} \bibnamefont{Chen}},
  \bibinfo{author}{\bibfnamefont{X.}~\bibnamefont{Dai}},
  \bibinfo{author}{\bibfnamefont{J.~L.} \bibnamefont{Luo}},
  \bibinfo{author}{\bibfnamefont{Z.}~\bibnamefont{Fang}}, \bibnamefont{and}
  \bibinfo{author}{\bibfnamefont{N.~L.} \bibnamefont{Wang}},
  \bibinfo{journal}{EPL} \textbf{\bibinfo{volume}{83}}, \bibinfo{pages}{27006}
  (\bibinfo{year}{2008}).

\bibitem[{\citenamefont{Chen et~al.}(2009)\citenamefont{Chen, Ren, Qiu, Bao,
  Liu, Wu, Wu, Xie, Wang, Huang, and Chen}}]{CRQB+08}
\bibinfo{author}{\bibfnamefont{H.}~\bibnamefont{Chen}},
  \bibinfo{author}{\bibfnamefont{Y.}~\bibnamefont{Ren}},
  \bibinfo{author}{\bibfnamefont{Y.}~\bibnamefont{Qiu}},
  \bibinfo{author}{\bibfnamefont{W.}~\bibnamefont{Bao}},
  \bibinfo{author}{\bibfnamefont{R.~H.} \bibnamefont{Liu}},
  \bibinfo{author}{\bibfnamefont{G.}~\bibnamefont{Wu}},
  \bibinfo{author}{\bibfnamefont{T.}~\bibnamefont{Wu}},
  \bibinfo{author}{\bibfnamefont{Y.~L.} \bibnamefont{Xie}},
  \bibinfo{author}{\bibfnamefont{X.~F.} \bibnamefont{Wang}},
  \bibinfo{author}{\bibfnamefont{Q.}~\bibnamefont{Huang}}, \bibnamefont{and}
  \bibinfo{author}{\bibfnamefont{X.~H.} \bibnamefont{Chen}},
  \bibinfo{journal}{EPL} \textbf{\bibinfo{volume}{85}}, \bibinfo{pages}{17006}
  (\bibinfo{year}{2009}).

\bibitem[{\citenamefont{Yin et~al.}(2008)\citenamefont{Yin, Lebegue, Han, Neal,
  Savrasov, and Pickett}}]{YLHN+08}
\bibinfo{author}{\bibfnamefont{Z.~P.} \bibnamefont{Yin}},
  \bibinfo{author}{\bibfnamefont{S.}~\bibnamefont{Lebegue}},
  \bibinfo{author}{\bibfnamefont{M.~J.} \bibnamefont{Han}},
  \bibinfo{author}{\bibfnamefont{B.~P.} \bibnamefont{Neal}},
  \bibinfo{author}{\bibfnamefont{S.~Y.} \bibnamefont{Savrasov}},
  \bibnamefont{and} \bibinfo{author}{\bibfnamefont{W.~E.}
  \bibnamefont{Pickett}}, \bibinfo{journal}{Phys. Rev. Lett.}
  \textbf{\bibinfo{volume}{101}}, \bibinfo{pages}{047001}
  (\bibinfo{year}{2008}).

\bibitem[{\citenamefont{Yildirim}(2008)}]{Yil08}
\bibinfo{author}{\bibfnamefont{T.}~\bibnamefont{Yildirim}},
  \bibinfo{journal}{Phys. Rev. Lett.} \textbf{\bibinfo{volume}{101}},
  \bibinfo{pages}{057010} (\bibinfo{year}{2008}).

\bibitem[{\citenamefont{Ishibashi et~al.}(2008)\citenamefont{Ishibashi,
  Terakura, and Hoson}}]{ITH08}
\bibinfo{author}{\bibfnamefont{S.}~\bibnamefont{Ishibashi}},
  \bibinfo{author}{\bibfnamefont{K.}~\bibnamefont{Terakura}}, \bibnamefont{and}
  \bibinfo{author}{\bibfnamefont{H.}~\bibnamefont{Hoson}}, \bibinfo{journal}{J.
  Phys. Soc. Jpn.} \textbf{\bibinfo{volume}{77}}, \bibinfo{pages}{053709}
  (\bibinfo{year}{2008}).

\bibitem[{\citenamefont{Mazin et~al.}(2008{\natexlab{a}})\citenamefont{Mazin,
  Singh, Johannes, and Du}}]{MSJD08}
\bibinfo{author}{\bibfnamefont{I.~I.} \bibnamefont{Mazin}},
  \bibinfo{author}{\bibfnamefont{D.~J.} \bibnamefont{Singh}},
  \bibinfo{author}{\bibfnamefont{M.~D.} \bibnamefont{Johannes}},
  \bibnamefont{and} \bibinfo{author}{\bibfnamefont{M.~H.} \bibnamefont{Du}},
  \bibinfo{journal}{Phys. Rev. Lett.} \textbf{\bibinfo{volume}{101}},
  \bibinfo{pages}{057003} (\bibinfo{year}{2008}{\natexlab{a}}).

\bibitem[{\citenamefont{Singh and Du}(2008)}]{SD08}
\bibinfo{author}{\bibfnamefont{D.~J.} \bibnamefont{Singh}} \bibnamefont{and}
  \bibinfo{author}{\bibfnamefont{M.-H.} \bibnamefont{Du}},
  \bibinfo{journal}{Phys. Rev. Lett.} \textbf{\bibinfo{volume}{100}},
  \bibinfo{pages}{237003} (\bibinfo{year}{2008}).

\bibitem[{\citenamefont{Boeri et~al.}(2008)\citenamefont{Boeri, Dolgov, and
  Golubov}}]{BDG08}
\bibinfo{author}{\bibfnamefont{L.}~\bibnamefont{Boeri}},
  \bibinfo{author}{\bibfnamefont{O.~V.} \bibnamefont{Dolgov}},
  \bibnamefont{and} \bibinfo{author}{\bibfnamefont{A.~A.}
  \bibnamefont{Golubov}}, \bibinfo{journal}{Phys. Rev. Lett.}
  \textbf{\bibinfo{volume}{101}}, \bibinfo{pages}{026403}
  (\bibinfo{year}{2008}).

\bibitem[{\citenamefont{Singh}(2008)}]{Singh08_Ba}
\bibinfo{author}{\bibfnamefont{D.~J.} \bibnamefont{Singh}},
  \bibinfo{journal}{Phys. Rev. B} \textbf{\bibinfo{volume}{78}},
  \bibinfo{pages}{094511} (\bibinfo{year}{2008}).

\bibitem[{\citenamefont{Opahle et~al.}(2009)\citenamefont{Opahle, Kandpal,
  Zhang, Gros, and Valenti}}]{OKZG+09}
\bibinfo{author}{\bibfnamefont{I.}~\bibnamefont{Opahle}},
  \bibinfo{author}{\bibfnamefont{H.~C.} \bibnamefont{Kandpal}},
  \bibinfo{author}{\bibfnamefont{Y.}~\bibnamefont{Zhang}},
  \bibinfo{author}{\bibfnamefont{C.}~\bibnamefont{Gros}}, \bibnamefont{and}
  \bibinfo{author}{\bibfnamefont{R.}~\bibnamefont{Valenti}},
  \bibinfo{journal}{Phys. Rev. B} \textbf{\bibinfo{volume}{79}},
  \bibinfo{pages}{024509} (\bibinfo{year}{2009}).

\bibitem[{\citenamefont{Nekrasov et~al.}(2008)\citenamefont{Nekrasov,
  Pchelkina, and Sadovskii}}]{NPS08_comp}
\bibinfo{author}{\bibfnamefont{I.~A.} \bibnamefont{Nekrasov}},
  \bibinfo{author}{\bibfnamefont{Z.~V.} \bibnamefont{Pchelkina}},
  \bibnamefont{and} \bibinfo{author}{\bibfnamefont{M.~V.}
  \bibnamefont{Sadovskii}}, \bibinfo{journal}{JETP Lett.}
  \textbf{\bibinfo{volume}{88}}, \bibinfo{pages}{144} (\bibinfo{year}{2008}).

\bibitem[{\citenamefont{Mazin et~al.}(2008{\natexlab{b}})\citenamefont{Mazin,
  Johannes, Boeri, Koepernik, and Singh}}]{MJBK+08}
\bibinfo{author}{\bibfnamefont{I.~I.} \bibnamefont{Mazin}},
  \bibinfo{author}{\bibfnamefont{M.~D.} \bibnamefont{Johannes}},
  \bibinfo{author}{\bibfnamefont{L.}~\bibnamefont{Boeri}},
  \bibinfo{author}{\bibfnamefont{K.}~\bibnamefont{Koepernik}},
  \bibnamefont{and} \bibinfo{author}{\bibfnamefont{D.~J.} \bibnamefont{Singh}},
  \bibinfo{journal}{Phys. Rev. B} \textbf{\bibinfo{volume}{78}},
  \bibinfo{pages}{085104} (\bibinfo{year}{2008}{\natexlab{b}}).

\bibitem[{\citenamefont{Andersen}(1975)}]{And75}
\bibinfo{author}{\bibfnamefont{O.~K.} \bibnamefont{Andersen}},
  \bibinfo{journal}{Phys. Rev. B} \textbf{\bibinfo{volume}{12}},
  \bibinfo{pages}{3060} (\bibinfo{year}{1975}).

\bibitem[{\citenamefont{Perdew and Wang}(1992)}]{PW92}
\bibinfo{author}{\bibfnamefont{J.~P.} \bibnamefont{Perdew}} \bibnamefont{and}
  \bibinfo{author}{\bibfnamefont{Y.}~\bibnamefont{Wang}},
  \bibinfo{journal}{Phys. Rev. B} \textbf{\bibinfo{volume}{45}},
  \bibinfo{pages}{13244} (\bibinfo{year}{1992}).

\bibitem[{\citenamefont{Antonov et~al.}(1995)\citenamefont{Antonov, Perlov,
  Shpak, and Yaresko}}]{APSY95}
\bibinfo{author}{\bibfnamefont{V.~N.} \bibnamefont{Antonov}},
  \bibinfo{author}{\bibfnamefont{A.~Y.} \bibnamefont{Perlov}},
  \bibinfo{author}{\bibfnamefont{A.~P.} \bibnamefont{Shpak}}, \bibnamefont{and}
  \bibinfo{author}{\bibfnamefont{A.~N.} \bibnamefont{Yaresko}},
  \bibinfo{journal}{J. Magn. Magn. Mater} \textbf{\bibinfo{volume}{146}},
  \bibinfo{pages}{205} (\bibinfo{year}{1995}).

\bibitem[{\citenamefont{Andersen and Boeri}(unpublished)}]{OKA&Lilia}
\bibinfo{author}{\bibfnamefont{O.~K.} \bibnamefont{Andersen}} \bibnamefont{and}
  \bibinfo{author}{\bibfnamefont{L.}~\bibnamefont{Boeri}}
  (\bibinfo{year}{unpublished}).

\bibitem[{\citenamefont{Lee}(unpublished)}]{Lee}
\bibinfo{author}{\bibfnamefont{P.}~\bibnamefont{Lee}}
  (\bibinfo{year}{unpublished}).

\bibitem[{\citenamefont{Coldea et~al.}(2008)\citenamefont{Coldea, Fletcher,
  Carrington, Analytis, Bangura, Chu, Erickson, Fisher, Hussey, and
  McDonald}}]{CFCA+08}
\bibinfo{author}{\bibfnamefont{A.~I.} \bibnamefont{Coldea}},
  \bibinfo{author}{\bibfnamefont{J.~D.} \bibnamefont{Fletcher}},
  \bibinfo{author}{\bibfnamefont{A.}~\bibnamefont{Carrington}},
  \bibinfo{author}{\bibfnamefont{J.~G.} \bibnamefont{Analytis}},
  \bibinfo{author}{\bibfnamefont{A.~F.} \bibnamefont{Bangura}},
  \bibinfo{author}{\bibfnamefont{J.~H.} \bibnamefont{Chu}},
  \bibinfo{author}{\bibfnamefont{A.~S.} \bibnamefont{Erickson}},
  \bibinfo{author}{\bibfnamefont{I.~R.} \bibnamefont{Fisher}},
  \bibinfo{author}{\bibfnamefont{N.~E.} \bibnamefont{Hussey}},
  \bibnamefont{and} \bibinfo{author}{\bibfnamefont{R.~D.}
  \bibnamefont{McDonald}}, \bibinfo{journal}{Phys. Rev. Lett.}
  \textbf{\bibinfo{volume}{101}}, \bibinfo{pages}{216402}
  (\bibinfo{year}{2008}).

\bibitem[{\citenamefont{Blacha et~al.}(2002)\citenamefont{Blacha, Schwarz,
  Madsen, Kvasnica, and Luitz}}]{WIEN2K}
\bibinfo{author}{\bibfnamefont{P.}~\bibnamefont{Blacha}},
  \bibinfo{author}{\bibfnamefont{K.}~\bibnamefont{Schwarz}},
  \bibinfo{author}{\bibfnamefont{G.~K.~H.} \bibnamefont{Madsen}},
  \bibinfo{author}{\bibfnamefont{D.}~\bibnamefont{Kvasnica}}, \bibnamefont{and}
  \bibinfo{author}{\bibfnamefont{J.}~\bibnamefont{Luitz}},
  \bibinfo{journal}{WIEN2K, An augmented planewave+local orbitals program for
  calculating crystal properties (Technische Universit\"{a}t Wien, Austria)}
  (\bibinfo{year}{2002}), \urlprefix\url{http://www.wien2k.at}.

\bibitem[{\citenamefont{Gunnarsson}(1976)}]{Gun76}
\bibinfo{author}{\bibfnamefont{O.}~\bibnamefont{Gunnarsson}},
  \bibinfo{journal}{J. Phys. F: Met. Phys.} \textbf{\bibinfo{volume}{6}},
  \bibinfo{pages}{587} (\bibinfo{year}{1976}).

\bibitem[{\citenamefont{Poulsen et~al.}(1976)\citenamefont{Poulsen, Kollar, and
  Andersen}}]{PKA76}
\bibinfo{author}{\bibfnamefont{U.~K.} \bibnamefont{Poulsen}},
  \bibinfo{author}{\bibfnamefont{J.}~\bibnamefont{Kollar}}, \bibnamefont{and}
  \bibinfo{author}{\bibfnamefont{O.~K.} \bibnamefont{Andersen}},
  \bibinfo{journal}{J. Phys. F: Met. Phys.} \textbf{\bibinfo{volume}{6}},
  \bibinfo{pages}{L241} (\bibinfo{year}{1976}).

\bibitem[{\citenamefont{Sandratskii}(1991)}]{San91a}
\bibinfo{author}{\bibfnamefont{L.~M.} \bibnamefont{Sandratskii}},
  \bibinfo{journal}{J. Phys.: Condens. Matter} \textbf{\bibinfo{volume}{3}},
  \bibinfo{pages}{8565} (\bibinfo{year}{1991}).

\bibitem[{\citenamefont{Sharma et~al.}(unpublished)\citenamefont{Sharma,
  Dewhurst, Shallcross, Bersier, Cricchio, Sanna, Massidda, Gross, and
  Nordstr\"om}}]{x:SDSB+08}
\bibinfo{author}{\bibfnamefont{S.}~\bibnamefont{Sharma}},
  \bibinfo{author}{\bibfnamefont{J.~K.} \bibnamefont{Dewhurst}},
  \bibinfo{author}{\bibfnamefont{S.}~\bibnamefont{Shallcross}},
  \bibinfo{author}{\bibfnamefont{C.}~\bibnamefont{Bersier}},
  \bibinfo{author}{\bibfnamefont{F.}~\bibnamefont{Cricchio}},
  \bibinfo{author}{\bibfnamefont{A.}~\bibnamefont{Sanna}},
  \bibinfo{author}{\bibfnamefont{S.}~\bibnamefont{Massidda}},
  \bibinfo{author}{\bibfnamefont{E.~K.~U.} \bibnamefont{Gross}},
  \bibnamefont{and}
  \bibinfo{author}{\bibfnamefont{L.}~\bibnamefont{Nordstr\"om}},
  \bibinfo{journal}{arXiv:0810.4278v2}  (\bibinfo{year}{unpublished}).

\bibitem[{\citenamefont{Callaway and Wang}(1975)}]{CW75}
\bibinfo{author}{\bibfnamefont{J.}~\bibnamefont{Callaway}} \bibnamefont{and}
  \bibinfo{author}{\bibfnamefont{C.}~\bibnamefont{Wang}}, \bibinfo{journal}{J.
  Phys. F: Met. Phys.} \textbf{\bibinfo{volume}{5}}, \bibinfo{pages}{2119}
  (\bibinfo{year}{1975}).

\bibitem[{\citenamefont{Yushankhai et~al.}(2007)\citenamefont{Yushankhai,
  Yaresko, Fulde, and Thalmeier}}]{YYFT07}
\bibinfo{author}{\bibfnamefont{V.}~\bibnamefont{Yushankhai}},
  \bibinfo{author}{\bibfnamefont{A.}~\bibnamefont{Yaresko}},
  \bibinfo{author}{\bibfnamefont{P.}~\bibnamefont{Fulde}}, \bibnamefont{and}
  \bibinfo{author}{\bibfnamefont{P.}~\bibnamefont{Thalmeier}},
  \bibinfo{journal}{Phys. Rev. B} \textbf{\bibinfo{volume}{76}},
  \bibinfo{pages}{085111} (\bibinfo{year}{2007}).

\end{thebibliography}

\end{document}